\RequirePackage{rotating}
\documentclass[usenatbib]{mnras}
\usepackage{hyperref}
\usepackage{epsfig}
\usepackage{amsmath}
\usepackage{graphicx}
\usepackage{array}
\usepackage{textcomp}
\usepackage{amssymb}
\usepackage{rotating}
\usepackage{pdflscape}
\usepackage{subcaption}
\usepackage{longtable}
\usepackage{xtab}
\def\mnras {{\sc MNRAS}}
\def\apj {ApJ}
\def\apjs {ApJS}
\def\apjl {ApJL}
\def\aj {AJ}

\def\pasp {PASP}

\def\aap {A\&A}

\title[S0s in MaNGA]
      {SDSS-IV MaNGA: The Formation Sequence of S0 Galaxies }
\author[A.\ Fraser-McKelvie et al.]
       {Amelia Fraser-McKelvie$^{1}$\thanks{Amelia.Fraser-McKelvie@nottingham.ac.uk}, 
Alfonso Arag\'on-Salamanca$^{1}$, Michael Merrifield$^{1}$, \and Martha Tabor$^{1}$, Mariangela Bernardi$^{2}$, Niv Drory$^{3}$, Taniya Parikh$^{4}$, \and Maria Argudo-Fern\'andez$^{5, 6}$.       
        \vspace*{1mm}\\
        $^{1}$ School of Physics \& Astronomy, University of Nottingham, University Park, Nottingham, NG7 2RD, U.K.\\
        $^{2}$ Department of Physics and Astronomy, University of Pennsylvania, Philadelphia, PA 19104, USA.\\
        $^{3}$ McDonald Observatory, The University of Texas at Austin, 1 University Station, Austin, TX 78712, USA.\\
        $^{4}$ Institute of Cosmology \& Gravitation, University of Portsmouth, Dennis Sciama Building, Portsmouth, PO1 3FX, U.K.\\
        $^{5}$ Centro de Astronom\'ia, Universidad de Antofagasta, Avenida Angamos 601, Antofagasta 1270300, Chile.\\
        $^{6}$ Chinese Academy of Sciences South America Center for Astronomy, China-Chile Joint Center for Astronomy, Camino El Observatorio, \\1515, Las Condes, Santiago, Chile.\\
	}
\begin{document}
\maketitle
\begin{abstract}
Gas stripping of spiral galaxies or mergers are thought to be the formation mechanisms of lenticular galaxies.
In order to determine the conditions in which each scenario dominates, we derive stellar populations of both the bulge and disk regions of 279 lenticular galaxies in the MaNGA survey.
We find a clear bimodality in stellar age and metallicity within the population of S0s and this is strongly correlated with stellar mass. Old and metal-rich bulges and disks belong to massive galaxies, and young and metal-poor bulges and disks are hosted by low-mass galaxies. From this we conclude that the bulges and disks are co-evolving. When the bulge and disk stellar ages are compared, we find that the bulge is almost always older than the disk for massive galaxies ($\textrm{M}_{\star} > 10^{10}~\textrm{M}_{\odot}$). The opposite is true for lower mass galaxies.
We conclude that we see two separate populations of lenticular galaxies. The old, massive, and metal-rich population possess bulges that are predominantly older than their disks, which we speculate may have been caused by morphological or inside-out quenching. In contrast, the less massive and more metal-poor population have bulges with more recent star formation than their disks. We postulate they may be undergoing bulge rejuvenation (or disk fading), or compaction. Environment doesn't play a distinct role in the properties of either population. Our findings give weight to the notion that while the faded spiral scenario likely formed low-mass S0s, other processes, such as mergers, may be responsible for high-mass S0s. 
\end{abstract}
\section{Introduction}
The lenticular (or S0) morphology class introduced by \citet{Hubble26} contains galaxies that consist of a bulge and a disk, but lack spiral arms.
A simple interpretation of the \citet{Dressler80} morphology-density relationship would imply that all lenticular galaxies were once spirals \citep{deV59,Sandage61, Dressler97,Fasano00,Desai07}. While all S0 galaxies appear morphologically similar, mounting evidence suggests that they encompass a wide range of formation processes and evolutionary pathways. Indeed, recent literature has revealed a range of photometric, spectroscopic, and kinematic properties in S0 galaxies \citep[e.g.][]{Laurikainen10, Barway13, Graham18}. From this, we conclude that lenticular galaxies may span a much wider range in galactic properties (and possibly formation pathways) than their relatively simple morphology would suggest.

The historical view that lenticulars form from faded spirals is supported by much prior literature \citep[e.g.][and references within]{BAM06, Moran07, Laurikainen10, Cappellari11, Prochaska11, Kormendy12, Johnston14}. In this scenario, gas within the spiral arms is either stripped through environmental mechanisms such as ram pressure stripping \citep{Gunn72}, harassment \citep{Moore98}, or strangulation \citep[e.g.][]{Larson80, Balogh00}, or the galaxy simply runs out of gas. What generally remains is a largely quiescent disk hosting a pseudo-bulge \citep{Kormendy04}.
This scenario has also been backed up with observations of spiral and S0 globular cluster number densities and specific frequencies \citep{Aragon-Salamanca06, Barr07}, and simulations \citep{Bekki11}. Given that lenticulars are more populous in denser environmental regions \citep{Dressler80, Postman84}, this scenario could be used to describe the formation of the majority of lenticulars.

The faded spiral paradigm cannot explain S0s whose properties vary significantly from expected progenitor spirals however \citep[e.g.][]{Williams10, Falcon-Barroso15}. 
In addition, recent literature has hinted at disparate bulge stellar populations within lenticulars \citep[e.g.][]{Johnston14, Tabor17, Mishra17}, often concluding that late-type spirals could not have plausibly evolved into the S0s we see today \citep[e.g.][]{Christlein04, Burnstein05,Gao18}.



A growing body of literature suggests that major mergers can also form S0 galaxies \citep[e.g.][]{Spitzer51,Tapia17, Diaz18, Eliche-Moral18, Mendez-Abreu18}. In this scenario, merger activity builds a bulge, and an inside-out formation scenario accretes gas onto a rotating disk. This sequence of events ensures that the bulge and disk possess separate formation histories, though simulations have also shown that it is possible to retain some degree of bulge and disk coupling throughout major merger events \citep{Querejeta15}.
Simulations also exist that create S0s by the fragmentation of an isolated, unstable cold disk over a short timescale \citep{Saha18}.

Internal secular evolution may also be involved in S0 formation \citep[e.g.][]{Laurikainen06}. Optically red spiral galaxies have bar fractions of 70-80\% \citep{Masters10, Fraser-McKelvie18}; these morphological features may be involved in the quenching of galaxies, funnelling gas towards the centre, inducing starbursts and bulge growth in these transition objects \citep[e.g.][]{Combes81,Pfenniger90}.

Whatever the formation mechanism, the overarching differentiator in lenticular structural relations seems to be luminosity \citep{Barway05, Aguerri12}. Many studies describe two separate populations of lenticular galaxies, one of which is luminous, with high bulge fractions and classical-type bulges, while the second fainter population contains less bulge-dominated galaxies and pseudo-bulges \citep[e.g.][]{Barway07, Barway09}. These trends hold irrespective of environment \citep{Barway05}. From this, it can be inferred that the rapid collapse scenarios required to produce classical bulges are produced by entirely different mechanisms to those creating the pseudo-bulges.
This might be explained by the fainter lenticulars being faded spiral descendants, and the more luminous S0s having had their stellar populations in place at high $z$, the classical bulge perhaps being created by repeated mergers \citep[e.g.][]{Barway13}.  
Furthermore, classical-bulge-hosting lenticulars tend to contain old stellar populations, while pseudo-bulge-hosting galaxies can still show signs of recent or even current star formation.

While this is a convenient argument, recent studies have placed pseudo-bulge-hosting lenticulars at both ends of the optical colour spectrum \citep{Mishra17}. 
The merger formation scenario may also not be required if classical bulge-hosting spirals evolve preferentially into S0s, as suggested by \citet{Mishra18}.
Irrespective of S0 formation mechanism, it seems that lenticulars cannot be separated by bulge type alone.  

 Stellar populations can vary dramatically across a galaxy, and it is becoming increasingly commonplace to analyse individual components of a galaxy separately \citep[e.g.][]{Johnston14, Rizzo17, Tabor17,Coccato18}. Bulge-disk decomposition is frequently employed as an analysis technique as it allows the separation of dispersion-dominated bulge light of a galaxy from the rotationally-supported disk regions. The advent of large-scale galaxy integral field spectroscopy (IFS) surveys such as Mapping Nearby Galaxies at APO \citep[MaNGA;][]{Bundy15} provides spatial information on thousands of nearby galaxies, from which stellar population information for various galaxy components may be derived.  
 
 This work aims to utilise the large sample of lenticular galaxies covering a wide mass range within the MaNGA survey.
We will probe the formation sequence of S0 galaxies by comparing the stellar populations of their bulge and disk regions. From this we will infer likely formation scenarios, and the dependence of different galaxy component evolutionary histories on one another. This paper is organised as follows: in Section~\ref{data} we describe the MaNGA survey and data products, along with our lenticular selection technique. In Section~\ref{results} we present line index and derived age and metallicity measurements for the bulge and disk regions of MaNGA lenticulars, then we discuss possible formation scenarios, comparing to the results of previous literature. In Section~\ref{conclusions}, we summarise our findings. 

\section{Data}
\label{data}
Lenticular galaxies were selected from the MaNGA survey's MaNGA product launch 5 (MPL-5), consisting of 2778 galaxy data cubes observed from March 2014 -- May 2016.
\subsection{The MaNGA Survey}
The MaNGA galaxy survey is an IFS survey that aims to observe $\sim$10,000 galaxies by 2020 \citep{Bundy15, Drory15} and is a project of SDSS-IV \citep{Blanton17} using the 2.5 metre telescope at the Apache Point Observatory \citep{Gunn06} and BOSS spectrographs \citep{Smee13}. 
The survey was designed to observe a luminosity-dependent sample of galaxies above a stellar mass of $10^{9}~\textrm{M}_{\odot}$, with a roughly flat $\log(\textrm{M}_{\star}$) distribution \citep{Law15, Wake17}. Because of the $\log(\textrm{M}_{\star}$) distribution criterion, more high-mass, red galaxies are observed than is expected for a representative sample of the local Universe. Each galaxy is therefore also assigned a weighting, which upweights or downweights a galaxy according to how under- or over-represented it is in the MaNGA sample. We may then consider the weighted sample to be volume-limited, and appropriate for statistical analysis.
A `Primary' sample of galaxies (50\% of targets) is defined with homogeneous spatial coverage out to $\sim1.5R_{e}$, (where $R_{e}$ is the Petrosian half-light radius in the $r$ band), with the addition of a `Colour-Enhanced' sample ($\sim$17\% of sample) to balance the colour distribution at fixed $\textrm{M}_{\star}$ \citep{Bundy15}. Combined together, the Primary and Colour-Enhanced samples, (termed the `Primary+' sample), make up $67\%$ of the MaNGA targets, the remaining $33\%$ of which comprise the `Secondary' sample of galaxies with extended coverage out to $\sim2.5R_{e}$.

All galaxy observations have wavelength coverage of $\sim$3500--10,000\AA, spectral resolution R $\sim2000$ (which gives an instrumental resolution of $\sim60$km/s), and an effective spatial resolution of 2.5$^{\prime\prime}$ (FWHM) after combining dithered observations; the sample lies within the redshift range $0.01<z<0.15$ \citep{Yan16b}. Observations are reduced by a data reduction pipeline \citep{Law16, Yan16} and made available as a single data cube per galaxy.

\subsection{Lenticular Galaxy Selection}
\begin{figure}
\centering
\includegraphics[width=0.50\textwidth]{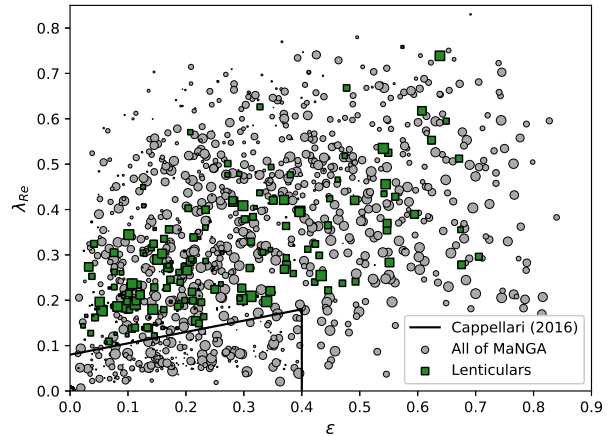} 
\caption{($\lambda_{Re}$, $\varepsilon$) diagram of both all of the galaxies in MPL-5 (grey points), and the regularly rotating lenticulars selected for this study (green). Data points are scaled by their volume weighting in the MPL-5 Primary+ sample. The slow rotator region, as defined by \citet{Cappellari16}, is denoted by a black line. The lenticulars in our sample are spread across the regularly-rotating area of ($\lambda_{Re}, \varepsilon$) space with no slow rotators, a consequence of our kinematic selection criterion.}
\label{lambdar_e}
\end{figure}

The traditional morphological definition of an S0 is a galaxy that contains a bulge and a disk, but lacks spiral arms. However, this definition can struggle to distinguish between S0s and ellipticals.
For this reason, we choose to employ kinematic selection techniques in tandem with a traditional morphological classification scheme.
We select lenticular galaxies from MPL-5 that have a Galaxy Zoo 2 \citep{Hart16} weighted `smooth' fraction $>0.7$. This is the fraction of respondents that determined a galaxy to be smooth and rounded, a selection aimed to separate featureless early-type galaxies from late-type spirals and barred galaxies. These values were also up- or down-weighted by the response record of the individual classifying, and corrected for redshift bias. This separation has the added benefit of removing barred S0s from our sample, which may contaminate bulge and disk photometric decomposition measurements.
We select only fast-rotating galaxies with: 
\begin{equation*}
\lambda_{Re} > 0.08 + \varepsilon_{e}/4  ~~~ \textrm{with} ~~~  \varepsilon_{e}>0.4,
\end{equation*}
where $\lambda_{Re}$ is the specific angular momentum, and $\varepsilon_{e}$ the ellipticity of the galaxy at 1 $R_{e}$ \citep{Cappellari16}. These selection criteria ensure we pick out rotating disks lacking spiral arm structure, while excluding edge-on galaxies in which spiral arms might be hidden. The smooth parameter also naturally selects against barred galaxies. From the 2778 galaxies in MPL-5, this procedure selects 279 lenticulars, the properties of which are summarised in Table~\ref{datatable}.
The sample spans the regular rotator parameter space on a  $\lambda_{Re}$ vs.~$\varepsilon$ diagram, as shown in Figure~\ref{lambdar_e}. 
The data points here are scaled by their volume weighting in the MPL-5 Primary+ sample: smaller data points represent galaxies that are over-represented in the MaNGA sample (for example, high-mass, optically-red galaxies), and large data points denote under-represented galaxies. 

\begin{figure}
\centering
\includegraphics[width=0.50\textwidth]{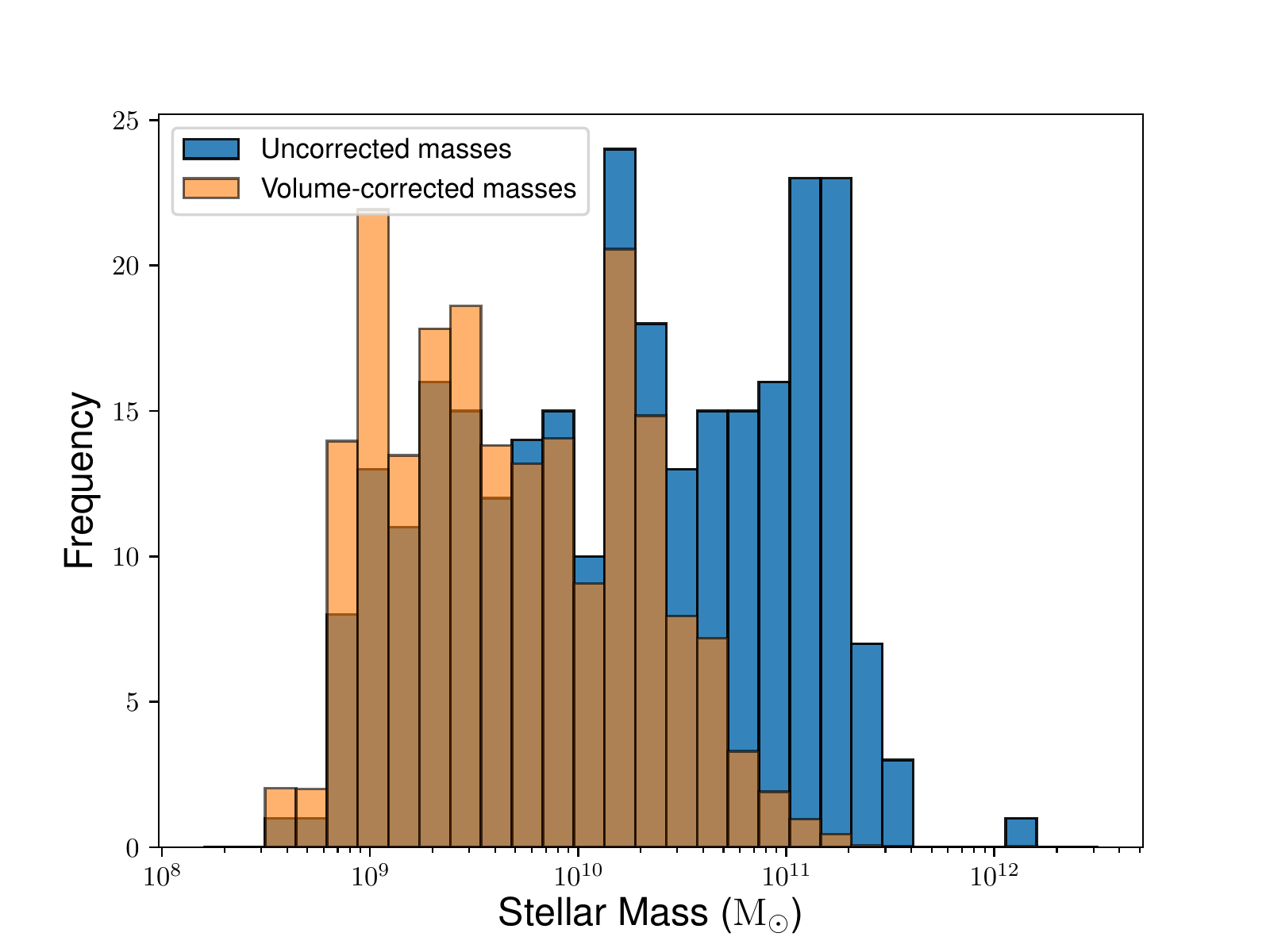}  
\caption{Histograms of the mass distribution of the lenticular sample before (blue) and after (orange) volume-weighting. 
The relatively flat $\log(\textrm{M}_{\star})$ distribution transforms to look more like a typical, local Universe mass function with the volume weightings provided for the Primary+ sample. }
\label{mass_dist}
\end{figure}

\begin{figure*}
\centering
\begin{subfigure}{0.8\textwidth}
\includegraphics[width=\textwidth]{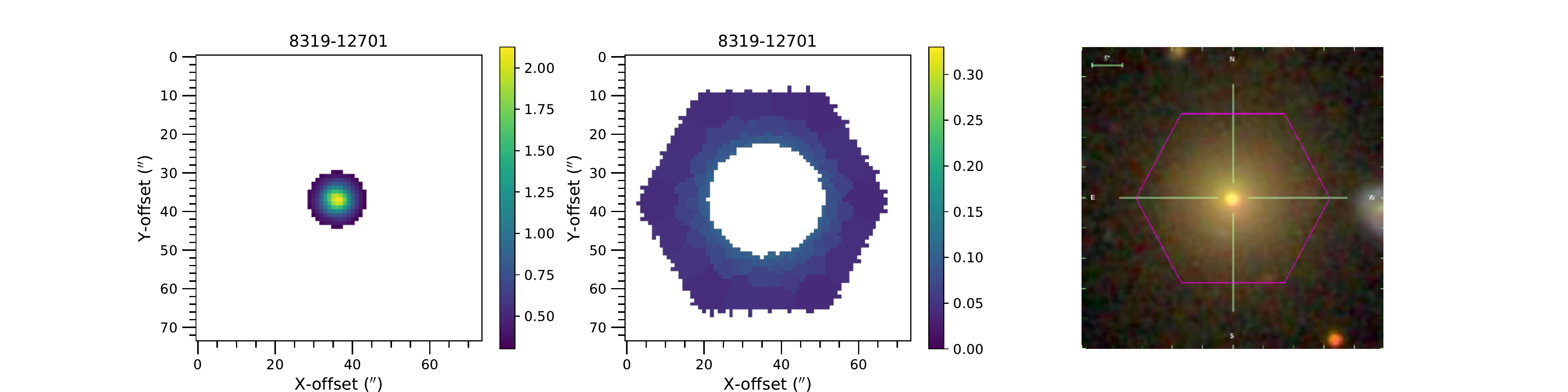} 
\end{subfigure}

\vskip\baselineskip

\begin{subfigure}{0.8\textwidth}
\includegraphics[width=\textwidth]{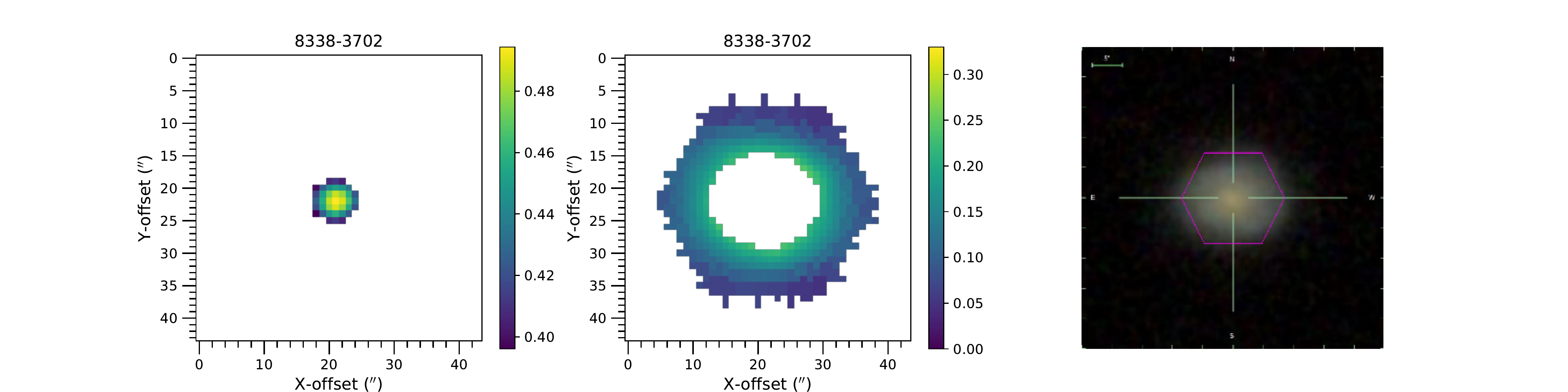}
\end{subfigure}
\caption{The bulge (left) and disk (centre) regions of two MaNGA lenticulars, 8319-12701 ($\textrm{M}_{\star}=9.6\times10^{10}~\textrm{M}_{\odot}$) and 8338-3702 ($\textrm{M}_{\star}=1.6\times10^{9}~\textrm{M}_{\odot}$). Bulge regions are defined as all spaxels within one $\textrm{R}_{\textrm{b}}$, and disk as spaxels outside 2$\textrm{R}_{\textrm{b}}$. The bulge and disk regions are coloured by the relative flux contribution of each Voronoi bin to the total flux of the galaxy. The right shows the colour images of the galaxy with the hexagonal MaNGA field of view overlaid in pink.}
\label{bulgedisk}
\end{figure*}

By construction, the volume weighting also has the effect of taking MaNGA's approximately flat $\log(\textrm{M}_{\star})$ distribution, and weighting the lower mass galaxies in such a way that the mass distribution begins to resemble the true mass function expected for the local Universe. Figure~\ref{mass_dist} shows a histogram of the mass distribution both before (blue) and after (orange) the volume weighting is applied to the sample. It should be noted that this sample spans a mass range of several dex. 

\subsection{Bulge and Disk Measurements}
In order to disentangle the effects of various galaxy components from one another, we separate the MaNGA data cubes into bulge and disk regions. We use the catalogue of \citet{Simard11}, who performed photometric bulge + disk decomposition on SDSS data release 7 galaxies. This catalogue provides fits to the images that allow the bulge S\'ersic index to vary freely based on the light profile of the galaxy, or constrain it such that $n_{\textrm{b}}=4$.
We check the F-test probabilities from the \citet{Simard11} catalogue to determine the best model to use for our data. When the $n_{\textrm{b}}=4$ bulge + disk and free $n_{\textrm{b}}$ bulge + disk models are compared, we find that the majority (66\%) of the lenticular sample are best fit using a free $n_{\textrm{b}}$ bulge + disk model, hence, this is what we use for our sample. 
 We have compared the $r$-band effective radius measurements for both fitting schemes and find general agreement in the photometric parameters of the fits over the sample. However, we note that the $n_{\textrm{b}}=4$ fits overestimate the bulge effective radius for galaxies where a more peaked bulge light distribution is favoured. Nevertheless, there is good agreement between the reported bulge to total ratios (B/T) for each technique. We therefore use the free $n_{\textrm{b}}$ models to characterise the bulge profiles in our sample. Bulge effective radius (R$_{\textrm{b}}$), $n_{\textrm{b}}$, B/T, and axis ratio (b/a) measurements are included in Table~\ref{datatable}.  

We define bulge regions to include all spaxels within one bulge effective radius of the central spaxel of a given galaxy, while disk regions are defined as spaxels outside two bulge effective radii and within the MaNGA field of view. Figure~\ref{bulgedisk} shows examples of the bulge and disk decompositions for two lenticulars in the sample, one high-mass, and one low-mass. 

\begin{figure}
\centering
\begin{subfigure}{0.45\textwidth}
\includegraphics[width=\textwidth]{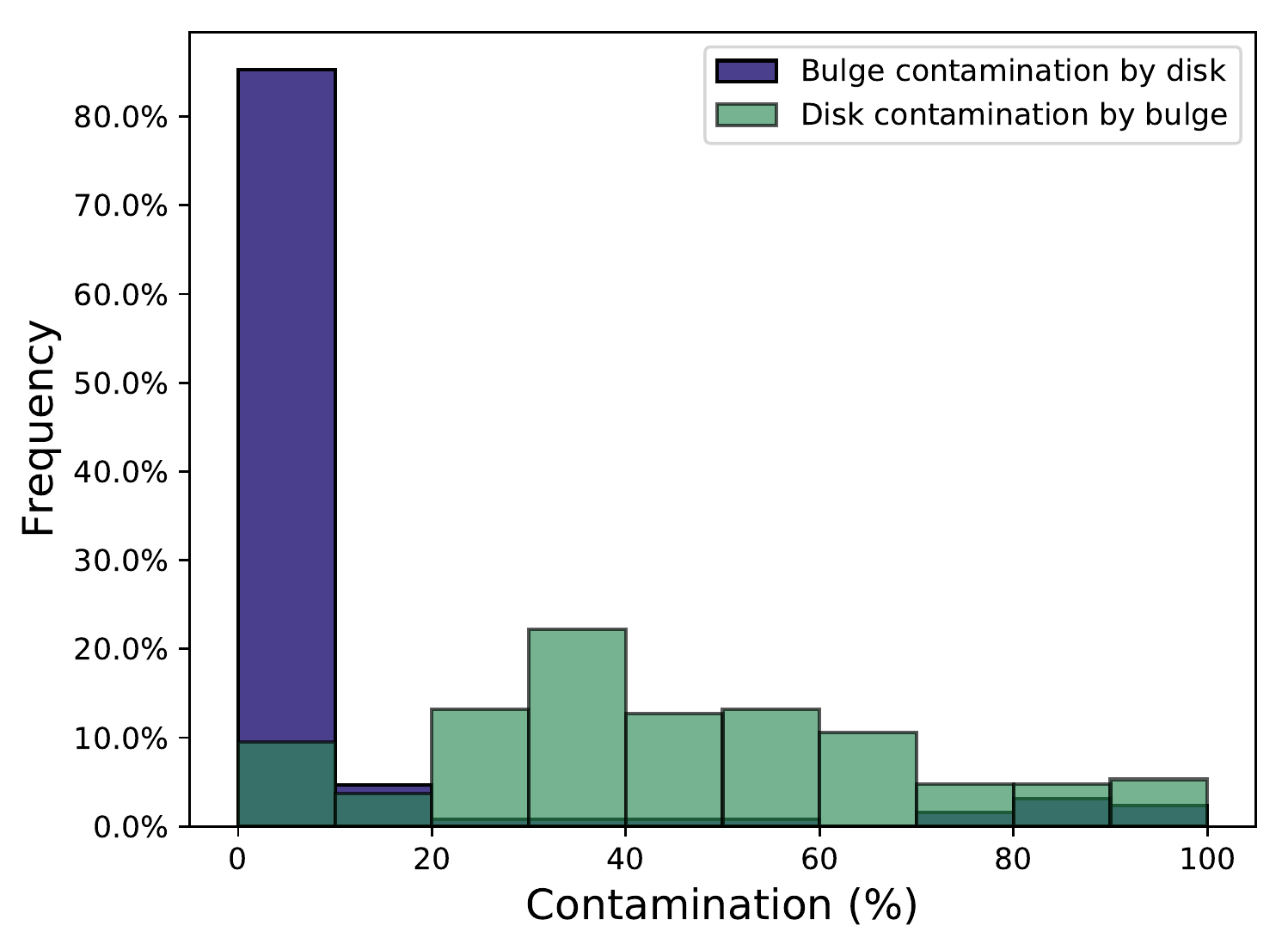}
\caption{}
\end{subfigure}

\hfill

\begin{subfigure}{0.45\textwidth}
\includegraphics[width=\textwidth]{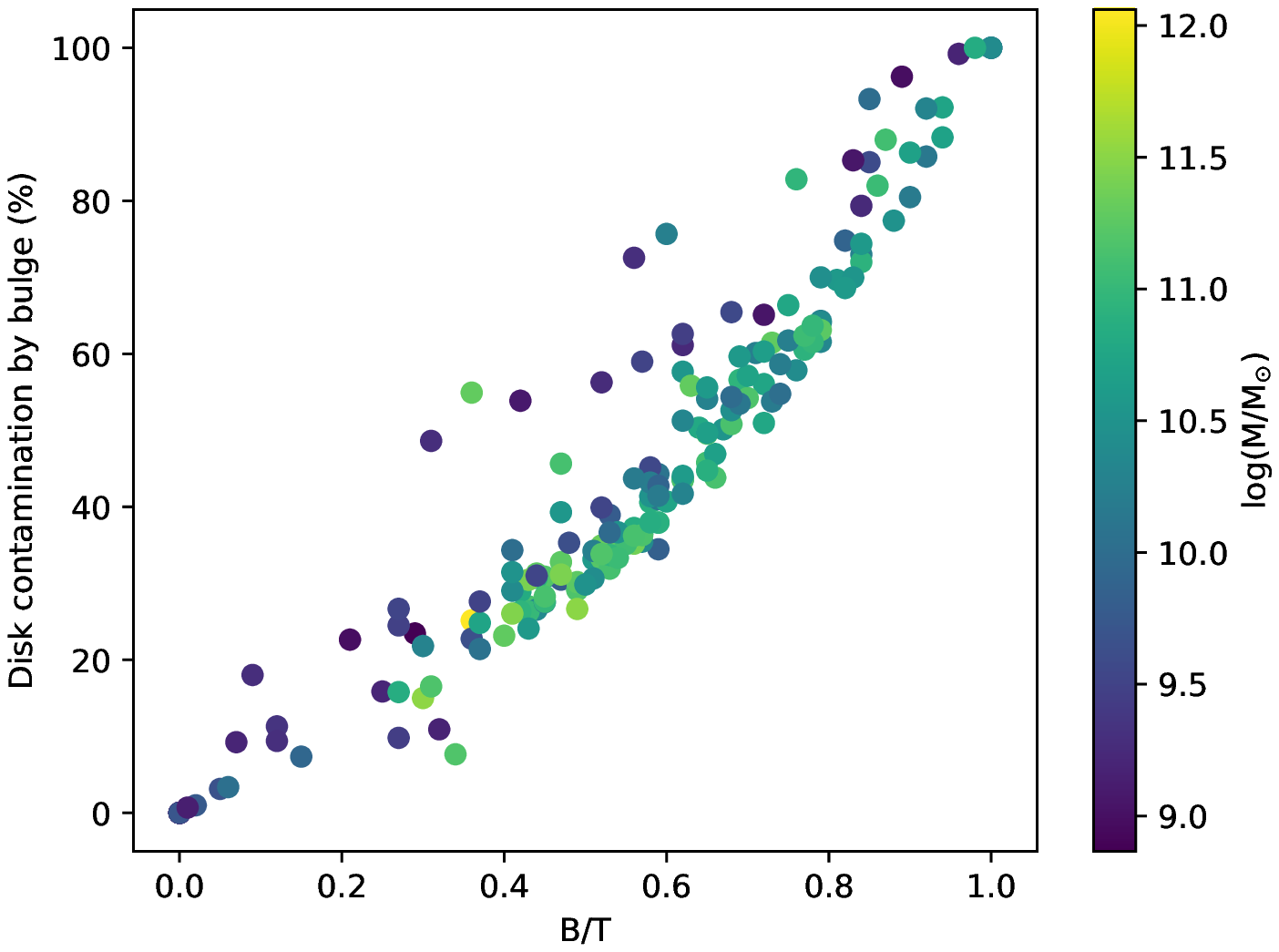}
\caption{}
\end{subfigure}
\caption{Quantification of the contamination of light within regions defined for each galaxy in the S0 sample. In panel a) we plot histograms of the fraction of contamination by the disk in the bulge regions (blue), and the bulge in the disk regions (green). While there is negligible contamination by the disk within bulge regions, some disks do have significant contamination of bulge light. In panel b) we plot the disk contamination by the bulge as a function of the B/T, with points coloured by stellar mass. As expected, contamination within the disk correlates with B/T.
We discuss the possible effects of this contamination on our results in the text.}
\label{contamination}
\end{figure}

In order to quantify any contamination occurring in either component as a result of the other, we calculate the fraction of disk light contaminating the bulge and bulge light contaminating the disk for every galaxy in our sample in Figure~\ref{contamination}a. We use the \citet{Simard11} values for $n_{\textrm{b}}$, R$_{\textrm{b}}$, and disk scale length ($h$) to compute bulge and disk light profiles, then calculate the fraction of bulge flux compared to total flux within one bulge effective radius for the bulge regions, and between two bulge effective radii and 1.5 $R_{e}$ for the disk regions. Rather than integrating to infinity, 1.5 $R_{e}$ was used as the edge of the disk as it is the target radius for the MaNGA Primary+ sample \citep{Bundy15}.

While there is negligible disk light contamination in all bulge regions, for some of the larger B/T galaxies we do observe a proportion of bulge light contributing to the disk region. In Figure~\ref{contamination}b, we plot the disk contamination by the bulge as a function of B/T, and colour data points by stellar mass. As expected, the fraction of disk contamination correlates with B/T. There are no systematic trends in B/T with stellar mass, though for a given B/T, the disks of lower-mass galaxies will generally be more contaminated than their higher mass counterparts.
We will investigate the effects of this bulge light contamination within disk regions on stellar population trends observed below.

For every galaxy in the sample, we measure the light-weighted average of the Lick indices H$\beta$, Mg$b$, Fe5270, and Fe5335 \citep[as defined by][]{Worthey94}, on MaNGA data which has been spatially Voronoi binned using the method of \citet{Cappellari03} to a signal to noise ratio of 10 per bin. 
We use the velocity dispersion-corrected index measurements provided by the MaNGA FIREFLY value-added catalog \citep{Abolfathi17}, which builds on the MaNGA data analysis pipeline (DAP; Westfall et al., in prep). These Lick indices have been corrected to MaNGA's spectral resolution using the stellar population models of \citet{Maraston11} and are measured from emission line subtracted spectra. 
The procedure for absorption line fitting is fully described in Section 2.3 of \citet{Parikh18}, but the basic recipe is to firstly fit and subtract emission lines from the spectrum using tools from the MaNGA DAP. The emission line-free spectrum is then used as input into the FIREFLY code, which determines the age and metallicity of the spectrum. The M11-MILES SSP combination closest to the age and metallicity derived from the full spectrum fitting from FIREFLY is then obtained. Then, the indices are measured on the model convolved to the wavelength-dependent SDSS resolution and convolved to the velocity dispersion of the spectrum as provided by the DAP. The ratio between these two measurements is the factor applied to the index measurements to correct them to the SDSS resolution.
We obtain the light-weighted average of these indices in all spaxels in the designated bulge and disk regions for all galaxies in the sample apart from MaNGA galaxy 8931-12705, which on closer inspection contains two separate galaxies in the field of view.

\section{Results \& Discussion}
\label{results}
\subsection{H$\beta$--[MgFe]$^{\prime}$ and Age--Metallicity Realtions}
\begin{figure*}
\centering
\begin{subfigure}{0.48\textwidth} 
\includegraphics[width=\textwidth]{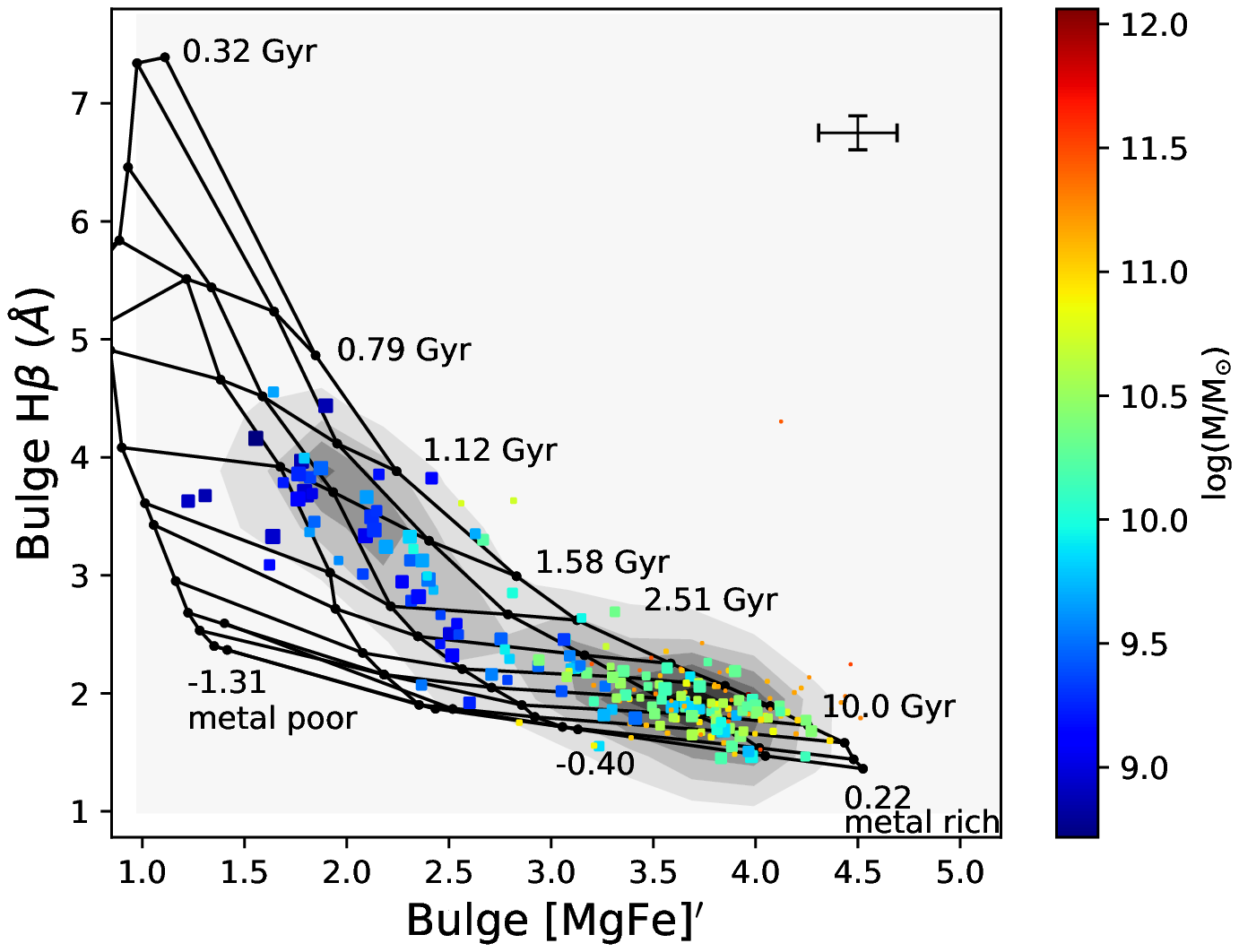}
\caption{Bulge}
\end{subfigure}
\hfill
\begin{subfigure}{0.48\textwidth} 
\includegraphics[width=\textwidth]{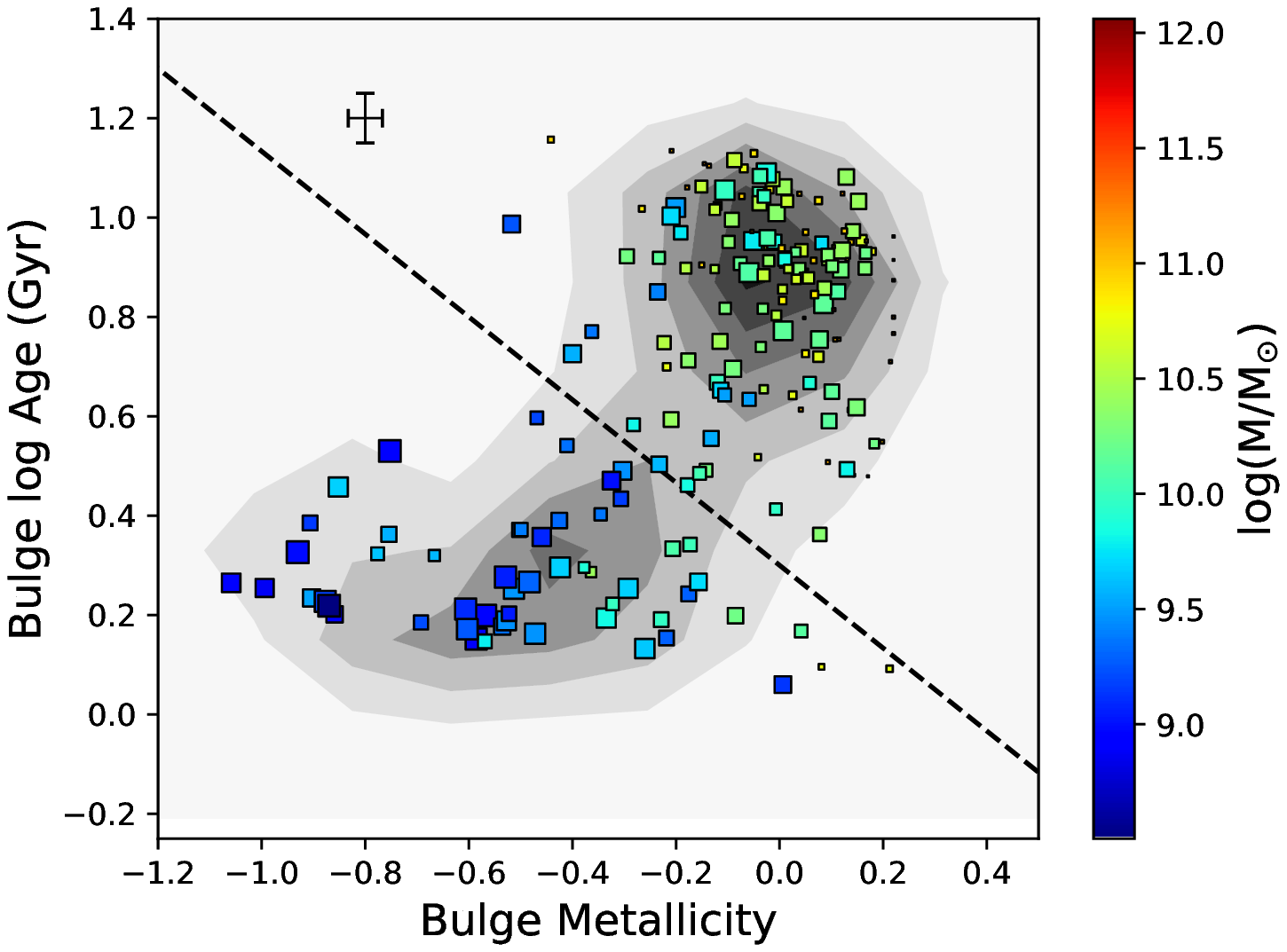}
\caption{Bulge}
\end{subfigure}

\vskip\baselineskip

\begin{subfigure}{0.48\textwidth}
\includegraphics[width=\textwidth]{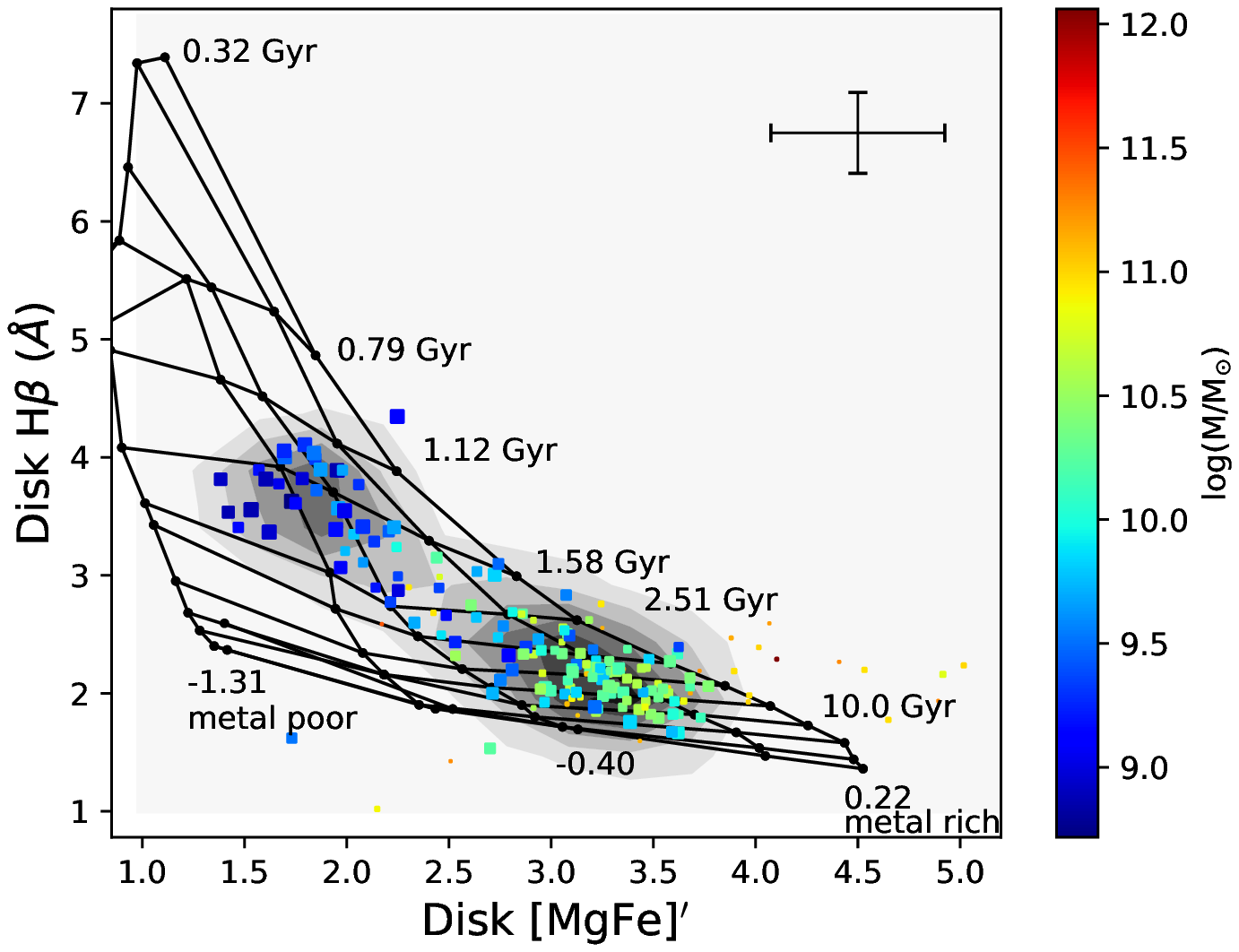}
\caption{Disk}
\end{subfigure}
\hfill
\begin{subfigure}{0.48\textwidth}
\includegraphics[width=\textwidth]{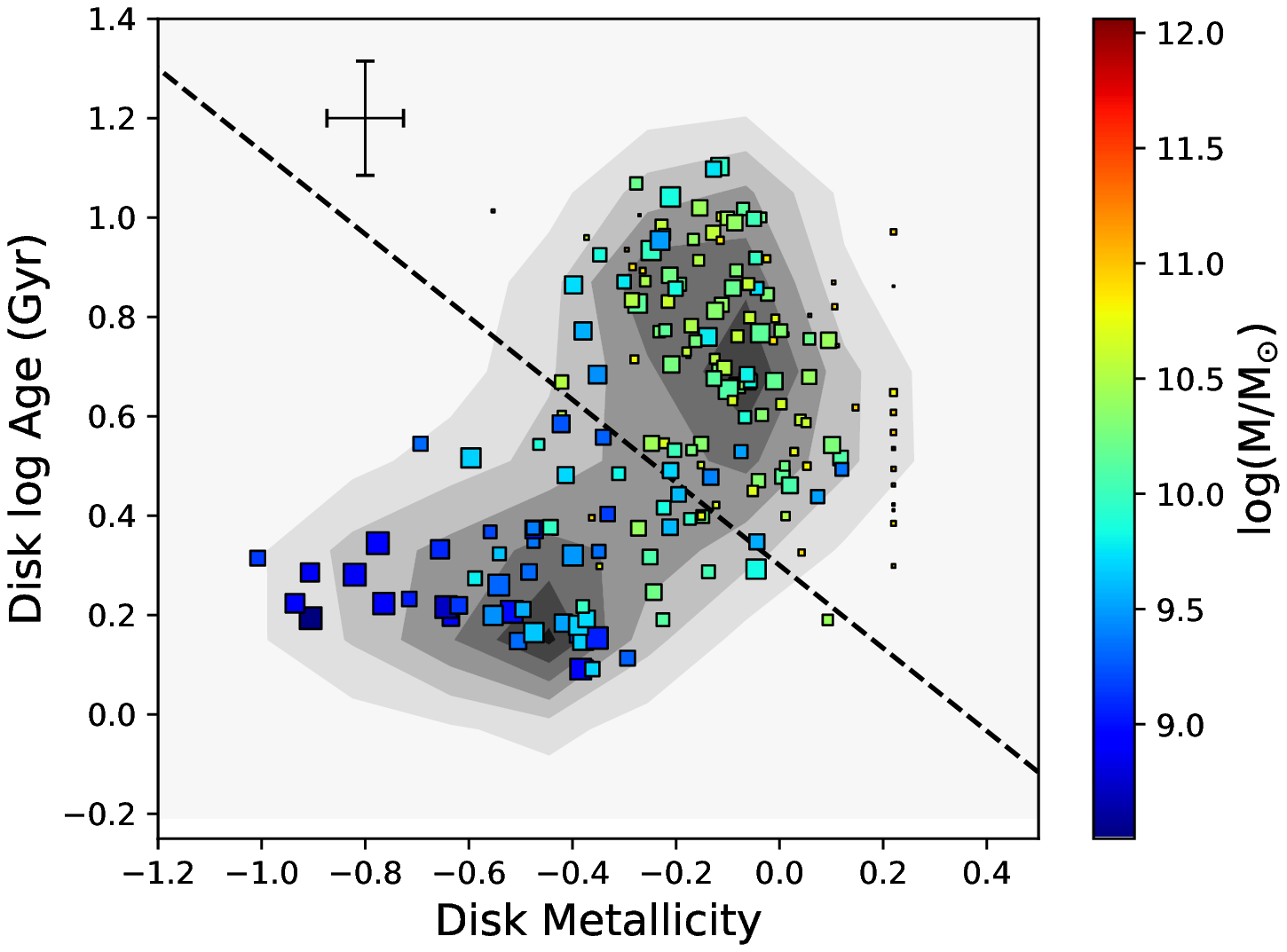}
\caption{Disk}
\end{subfigure}

\caption{Observables and their implied physical stellar ages and metallicities for the bulge (panels a and b) and disk (panels c and d) regions of MaNGA lenticulars.
The index--index diagram for the bulge regions is shown in panel a, and disk regions in panel c, with SSP model predictions of \citet{Vazdekis10} also overlaid.
These model lines are bi-linearly interpolated between to obtain stellar age and metallicity estimates for the bulge (panel b) and disk (panel d) regions. An arbitrary dashed line separates the two regions of more metal-rich, older stellar populations from the metal-poor, younger populations.
For all plots, the data points and contours are weighted by their volume weighting in the Primary+ sample, and points are colour-coded by their NSA stellar mass. Representative error bars derived from Monte Carlo errors on the Lick index measurements are also displayed. The contours show a clear bimodality in H$\beta$ and [MgFe]$^{\prime}$, and age and metallicity for both the bulge and disk regions of the MaNGA lenticulars. This trend correlates strongly with stellar mass.}
\label{Hb_MgFe}
\end{figure*}

We employ hydrogen, and magnesium and iron absorption line indices as indicators of stellar age and metallicity, respectively. 
In Figure~\ref{Hb_MgFe} we plot H$\beta$ vs. [MgFe]$^{\prime}$ line index measurements for the bulge (panel a) and disk (panel c) regions of the lenticulars in the sample. Here we adopt 
\begin{equation*}
[\textrm{MgFe}]^{\prime} = \sqrt{\textrm{Mg}b~(0.72 \times \textrm{Fe}5270 + 0.28 \times \textrm{Fe}5335)},
\end{equation*}
as this ratio is reasonably insensitive to changes in $\alpha/\textrm{Fe}$ ratio \citep{Gonzalez93, Thomas03}.

\begin{figure*}
\includegraphics[width=0.99\textwidth]{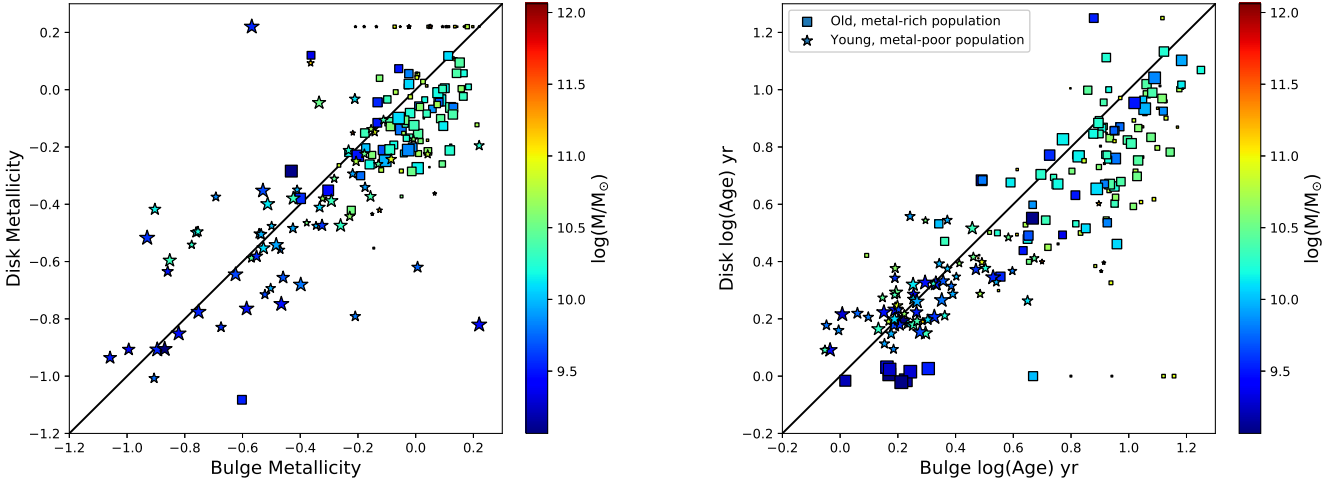}
\caption{Comparison of bulge and disk metallicities (left) and ages (right) for MaNGA lenticulars. Each point is coloured by its stellar mass and weighted by its volume weighting in the Primary+ sample, and galaxies that are located above the dividing line of Figure~\ref{Hb_MgFe}b (predominantly old and metal-rich) are shown with square markers. Those below the dividing line are shown in star markers.  In both panels there is a 1:1 line for comparison, and we see that on average, bulge and disk ages and metallicities are similar within a given galaxy. The bulges of higher-mass S0s are systematically older and more metal-rich than their disks. }
\label{ageage_metmet}
\end{figure*}

\begin{figure*}
\includegraphics[width=0.8\textwidth]{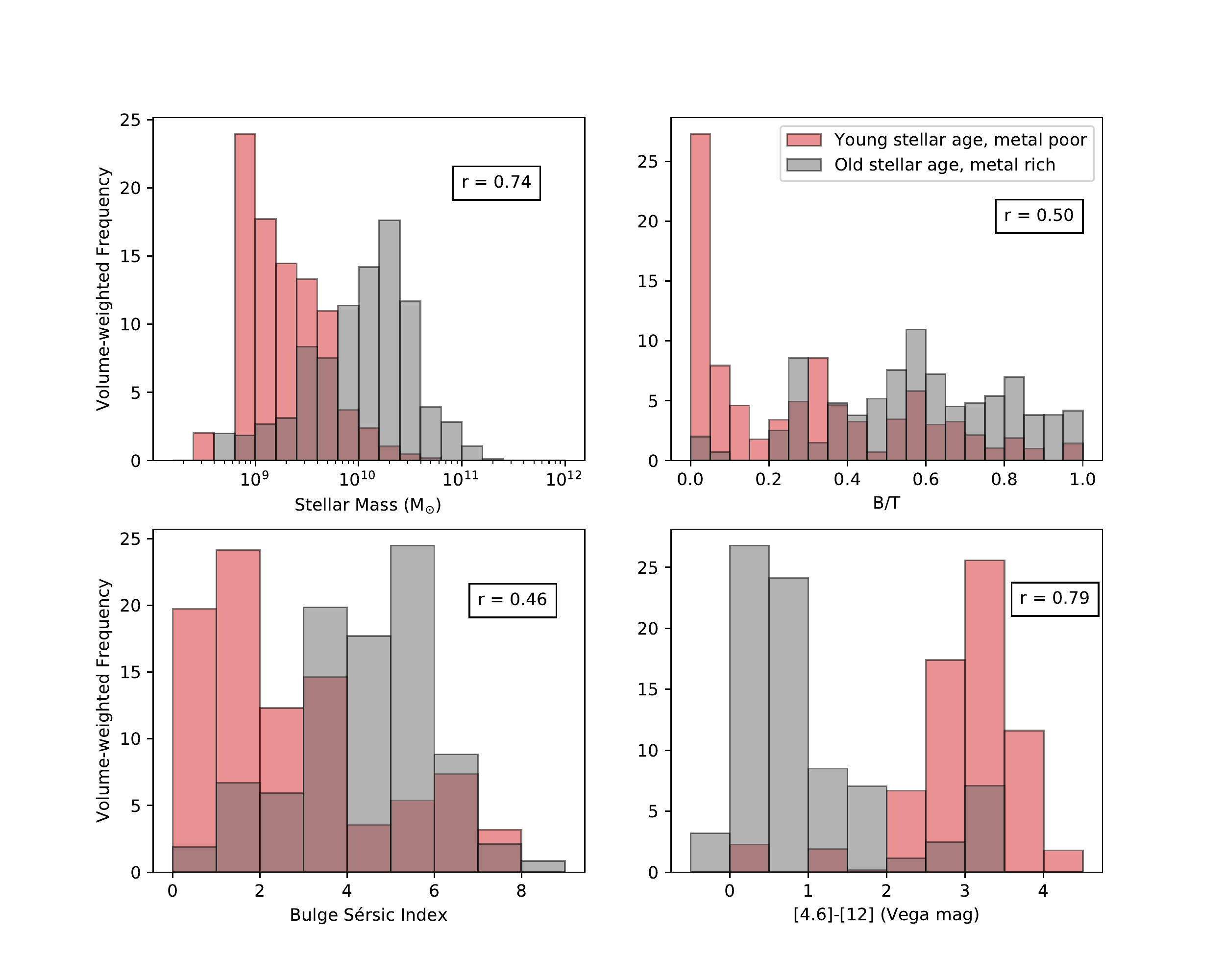} 
\caption{Histograms of physical properties of both the older, more metal-rich sample of lenticulars (grey), and the younger, more metal-poor sample (red). These populations are defined by considering the bulge age and metallicity with respect to the dividing line of Figure~\ref{Hb_MgFe}b. From top left: stellar masses, B/T, bulge S\'ersic index, and $WISE$ 4.6 $\mu$m--12$\mu$m colour. All properties show a somewhat bimodal distribution, with the exception perhaps of B/T. Correlation coefficients, r, between each quantity and the distance from each galaxy to the arbitrary separating line drawn in Figure~\ref{Hb_MgFe}b are also printed. $WISE$ 4.6 $\mu$m--12$\mu$m and stellar mass possess the strongest correlations.}
\label{massBThistograms}
\end{figure*}

Overlaid on the index--index plots are the single stellar population (SSP) model predictions of \citet{Vazdekis10} using Padova 2000 isochrones and a \citet{Chabrier03} initial mass function. Data points are scaled by their volume weighting in the Primary+ sample, and coloured by their stellar mass, which is taken from the NASA Sloan Atlas (NSA)\footnote{\url{http://www.nsatlas.org/}}, calculated using the photometric sky subtraction technique of \citet{Blanton11}. Monte Carlo-based errors on the line measurements were derived by perturbing the flux at each wavelength using the error in the flux derived by the MaNGA DAP for 100 realisations. The weighted contours show a clear bimodality in H$\beta$ and [MgFe]$^{\prime}$ for both the bulge and disk regions, and are well correlated with stellar mass. The bulge and disk H$\beta$, Mg$b$, Fe5270 and Fe5335 measurements for each galaxy are listed in Table~\ref{datatable2}.

To obtain estimates of bulge and disk stellar age and metallicity, we translate these measurements into physical properties by bi-linearly interpolating between the quadrilaterals formed by the SSP model lines. Where any data point lay outside of the model grid lines, we assigned it the age or metallicity of the nearest grid line. We note that while this method is a simplification, especially when compared to full spectral fitting models, it is closer to the actual data. Given the difficulties in obtaining robust star formation histories, along with their degeneracies and non-uniqueness \citep[see][for a summary]{Wilkinson17}, this simple method is more robust for relative comparisons between derived ages and metallicities.

In Figure~\ref{Hb_MgFe}, we plot the bulge (panel b) and disk (panel d) derived stellar ages and metallicities, again with data points scaled by their volume weighting in the MaNGA Primary+ sample and coloured by stellar mass. We see two clear populations of lenticulars -- one that is metal-rich with an old stellar population and high stellar masses, and the other that comprises a younger stellar population, and is more metal-poor. We separate these populations by an arbitrary dashed line, chosen to distinguish between the two separate populations as well as possible. Trends in bulge and disk age and metallicity are more clearly seen in Figure~\ref{ageage_metmet}, where bulge metallicity is plotted against disk metallicity, and bulge age against disk age. Galaxies that are old, metal-rich with high stellar mass that are located above the dividing line of Figure~\ref{Hb_MgFe} are shown with square markers, and those below the dividing line with star-shaped markers.
Again, we see the bulges of higher-mass S0s are almost always older and more metal-rich than their disks. Given that from Figure~\ref{contamination}b we see the fraction of disk contamination by the bulge seems to be insensitive to stellar mass, we confirm that any trends seen are not an effect of bulge contamination as a function of mass.


Interestingly, in Figure~\ref{ageage_metmet}, we see the high-mass galaxies that are dominated by metal-rich, old stellar populations in their core also possess similar stellar populations in their disk (albeit slightly younger and more metal-poor). High-mass lenticulars almost exclusively host older and more metal-rich bulges and disks than their lower mass counterparts. We conclude from this that the bulges and disks of lenticulars, at least to some degree, are co-evolving \citep[e.g.][]{Balcells94, Laurikainen10}. 

When we divide the two populations by the arbitrary dashed line shown in Figure~\ref{Hb_MgFe}b and d, the separation in mass between these two populations corresponds to lenticulars of stellar mass $\lesssim 10^{10}~\textrm{M}_{\odot}$. We note this is a similar mass to that reported by \citet{Tasca09} ($\sim6\times10^{10}~\textrm{M}_{\odot}$), above which spirals are less likely to evolve, irrespective of their environment. If spirals are less likely to evolve, this may be further evidence that high-mass lenticulars formed from mechanisms other than spiral disk fading, for example, dissipational processes. 

We investigate trends in physical properties for these two populations of lenticulars in Figure~\ref{massBThistograms} by separating the lenticular population into two sub-samples, defined by which side of the dividing line of Figure~\ref{Hb_MgFe}b they inhabit. While there is a clear division in stellar mass between the older, metal-rich population and the younger, metal-poor lenticulars, we also see strong correlations in bulge S\'{e}rsic index and Wide-Field Survey Explorer ($WISE$) mid-infrared (mid-IR) colour. Bulge S\'ersic indices are obtained from the \citet{Simard11} catalogue, and we see that a large proportion of young metal-poor lenticulars possess bulges of S\'ersic index $n_{b}<2$, and nearly all old, metal-rich lenticulars have $n_{b}>2$. This suggests that pseudo-bulges may be more prevalent in the young, metal-poor population, and classical bulges in the older, metal-rich population, which in turn, adds further support to the separate formation mechanisms scenario \citep[e.g.][]{Fisher08, Fisher16}. The correlation with $n_{b}$ is weaker than with stellar mass, shown by the correlation coefficient, $r$, between each property and the distance of each galaxy to the dividing line, shown in Figure~\ref{massBThistograms}, where a coefficient of $r = 1$ corresponds to a perfect correlation between two quantities.

$WISE$ 4.3 $\mu$m -- 12 $\mu$m colour is an excellent indicator of recent star formation activity \citep{Jarrett11, Cluver14}, and we see a very obvious division in this quantity between the two lenticular populations in Figure~\ref{massBThistograms}. The more passive mid-IR colours of the older and more metal rich population indicate they have likely stopped forming stars, while the younger, more metal-poor population possess mid-IR colours consistent with recent star formation, possibly through a rejuvenation scenario. The only physical quantity that does not possess a strong bimodality is B/T. There are many galaxies that are metal-poor with younger stellar populations but large B/T. From this, we infer that it is stellar mass, and not B/T, that drives the division in lenticular stellar populations.

The striking bimodality in both stellar age and metallicity by stellar mass (and, presumably, luminosity) agrees well with the work of \citet{Barway07,Barway09}, who propose that less luminous lenticulars have been formed by secular processes gently stripping gas from spiral disks, and the more luminous S0s have been formed at high redshift by violent, dissipative processes.  
Fainter samples such as those presented in \citet{BAM06} still have young and metal-poor stellar populations, which we would expect from disks possessing pseudo-bulges. 
Works that study only high-mass lenticulars, such as \citet{Mendez-Abreu18}, conclude that high-mass lenticulars were likely formed by dissipational processes such as major mergers or gas accretion at high redshift.
From this it is possible to conclude that the formation pathway taken by a lenticular may chiefly be determined by its mass, and in this study which spans a wide range of masses we see clear evidence of these multiple pathways.

\begin{figure*}
\centering
\begin{subfigure}{0.33\textwidth} 
\includegraphics[width=\textwidth]{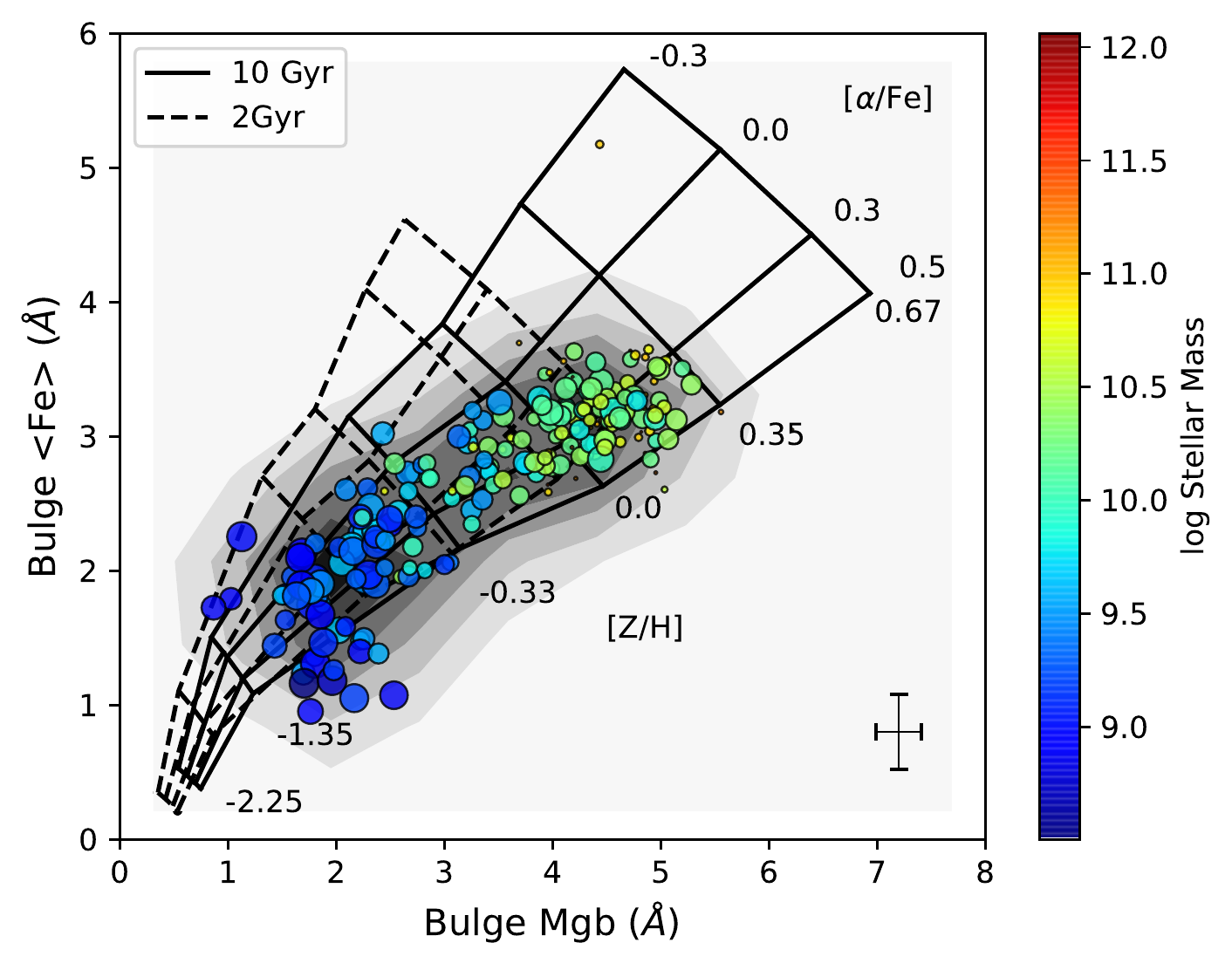}
\caption{Bulge}
\end{subfigure}
\hfill
\begin{subfigure}{0.33\textwidth} 
\includegraphics[width=\textwidth]{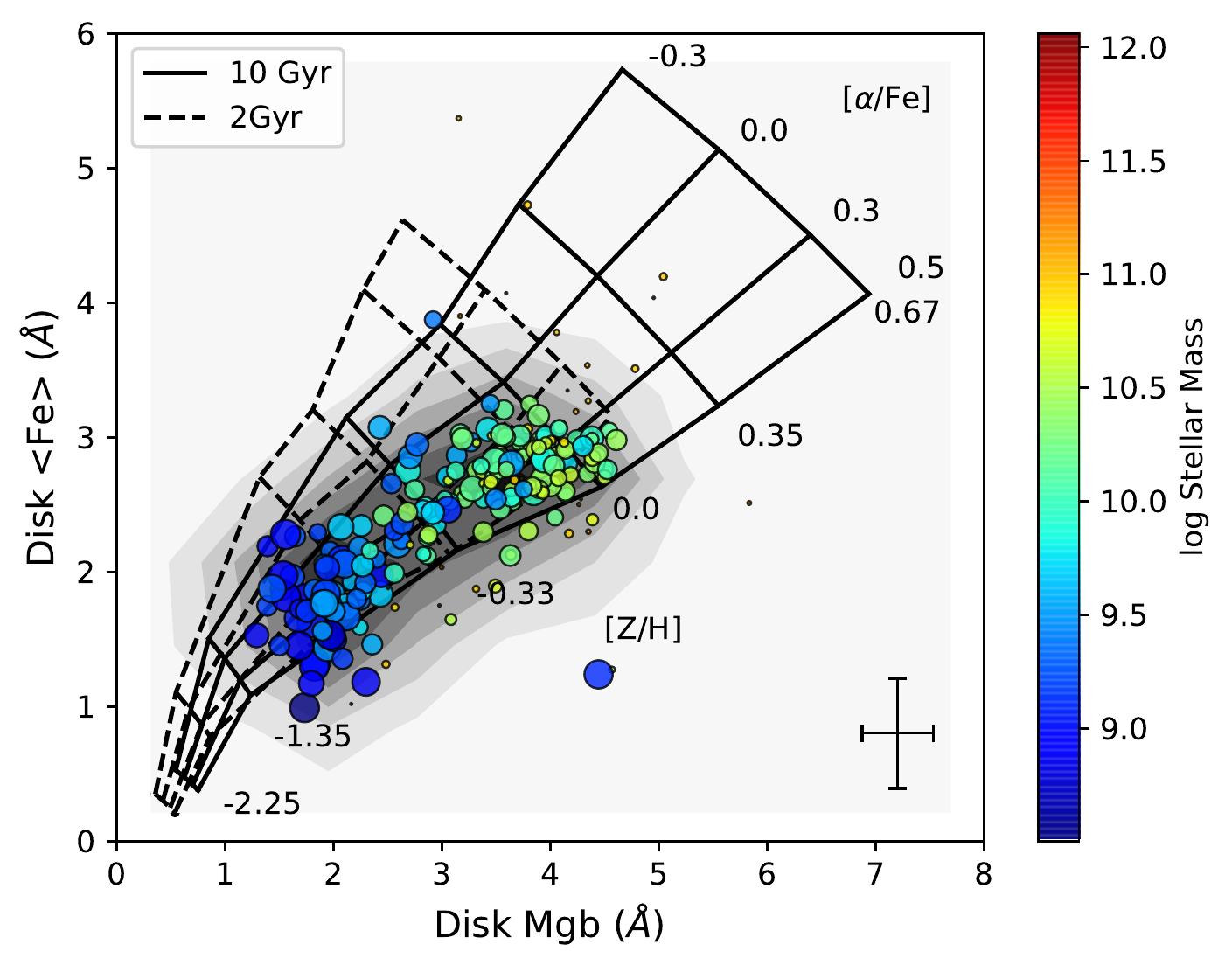}
\caption{Disk}
\end{subfigure}
\hfill
\begin{subfigure}{0.33\textwidth}
\includegraphics[width=\textwidth]{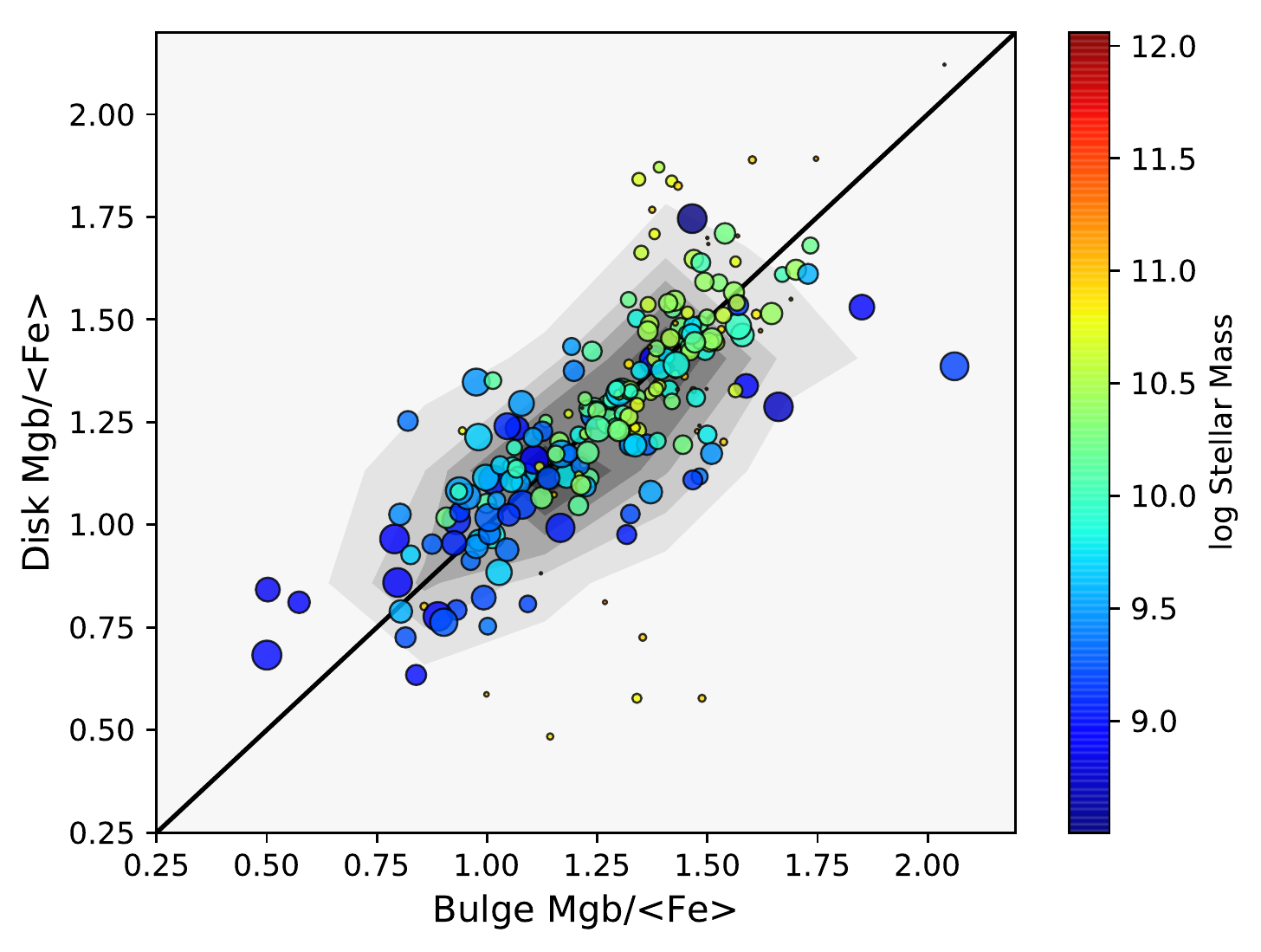} 
\caption{Bulge vs. disk}
\end{subfigure}
\caption{Investigation into the $\alpha$-enhancement of the S0 population. Bulge (panel a) and disk (panel b) Mg$b$ and $\langle \textrm{Fe} \rangle$ measurements are presented with \citet{Thomas11} high resolution SSP models overlaid for 10 Gyr (solid) and 2 Gyr (dashed) models.  The data points are again weighted by their volume weighting in the Primary+ sample, and points are colour-coded by their stellar mass with representative error bars in the corner. The average [$\alpha$/Fe] is $\sim$0.3 for the entire S0 sample in both the bulge and disk regions, and observed trends in the Mg$b$ and Fe indices are hence the result of metallicity differences between the bulge and disk. 
In panel c we plot bulge vs. disk Mgb/$\langle \textrm{Fe} \rangle$ and see the $\alpha$-enhancement for bulge and disk is very similar at all masses.}
\label{met_alpha}
\end{figure*}

\subsection{$\langle \textrm{Fe} \rangle$--Mg$b$ and $\alpha$-element abundances}
The Lick indices Mg$b$ and $\langle \textrm{Fe} \rangle$, (where $\langle \textrm{Fe} \rangle$ is defined as $\langle \textrm{Fe} \rangle = (\textrm{Fe}5270 + \textrm{Fe}5335)/2$), are a good probe of $\alpha$-element abundances, which indicate the timescale of star formation. 
Type II supernovae eject $\alpha$-element-rich material shortly after star formation has commenced. In contrast, Type Ia supernovae are the product of older stars, and contribute only small amounts of $\alpha$-elements when they explode. Instead, they tend to enrich the interstellar medium with Fe.
Hence, massive early-type galaxies typically exhibit enhanced $\alpha$-abundance ratios \citep[e.g.][]{Thomas05}, thought to be a result of AGN feedback suppressing star formation at early times \citep[e.g.][]{Segers16}. Without this early truncation of star formation, lower-mass galaxies may retain star formation over longer timescales and host more Type Ia supernovae, which we should observe as a lower $\alpha$-enhancement.

In Figure~\ref{met_alpha} we plot the bulge and disk $\langle \textrm{Fe} \rangle$ and Mg$b$ indices for the sample of lenticular galaxies, again colour-coded by stellar mass with point size and contours weighted by a galaxy's over- or under-representation in the MaNGA sample. These values are also recorded in Table~\ref{datatable2}. The high-resolution stellar population models of \citet{Thomas11} are overlaid for a model age of 10 Gyr (solid lines) and 2 Gyr (dashed lines). While differences in bulge and disk metallicity are seen in panels a and b, there seems to be little difference in $\alpha$-abundance between the bulge and the disk. Panel c confirms this, as we plot bulge and disk Mg$b$/$\langle \textrm{Fe} \rangle$. The similarity in bulge and disk $\alpha$-enhancement regardless of galaxy mass or B/T leads us to believe that bulge contamination is not affecting the $\alpha$-enhancement measurements of the lenticular sample. 

The majority of lenticulars possess bulge and disk $\alpha$-enhancement above Solar values, with an average [$\alpha$/Fe]$\sim$0.3. If the Milky Way is considered a typical spiral galaxy, then the timescale for star formation for nearly all S0 galaxies is shorter than for spirals. 
This trend in $\alpha$-enhancement is seen regardless of S0 stellar mass or implied formation mechanism; S0s all quenched similarly quicker than the Milky Way.
We observe the $\alpha$-enhancement for the low-mass galaxies is more spread than that of the high-mass galaxies, indicating that star formation timescales are more varied for low-mass galaxies. We infer that low-mass galaxies are more fragile than their high-mass counterparts, and may be more significantly impacted by environmental trends \citep[e.g.][]{Maltby10, Fraser-McKelvie18}, providing a wider variety of star formation histories. 

\citet{Johnston14} found that the disks of Virgo cluster S0s (all with NSA stellar masses $<~4\times10^{10}~\textrm{M}_{\odot}$) were slightly more $\alpha$-enhanced than the bulges. This implies the star formation timescales of disks are shorter than within the bulges; they presented a scenario wherein the gas that produced the final star formation event in the bulge was pre-enriched by earlier star formation in the disk. They suggest this may have occurred as gas was dumped into the core of a galaxy after a stripping event. This same event may have quenched the disk and transformed spiral morphology to lenticular. We do not observe this trend, and also note that we have no cluster S0s in our sample. It is possible this effect is a cluster-only phenomenon, and further work on lenticulars with a wide range of masses living in varied environments is required. 
\subsection{Environment}
In order to determine whether galaxy large-scale environment is having an effect on stellar populations within S0 galaxies, we present two different environmental indicators: the tidal strength parameter, $Q_{\textrm{lss}}$, and the projected galaxy number density, $\eta_{k}$. These quantities are defined in \citet{Argudo-Fernandez15}, and briefly summarised here. The tidal strength parameter, $Q_{\textrm{lss}}$, is an estimation of the local tidal strength at 1Mpc, and gives an indication of how strongly nearby galaxies are perturbing the galaxy in question. The greater the value of $Q_{\textrm{lss}}$, the less isolated the galaxy is from external influence. 
 $\eta_{k}$ is the projected density to the $5^{\textrm{th}}$ nearest neighbour. It is well correlated with the distance to the $5^{\textrm{th}}$ nearest neighbour, such that galaxies with a greater distance to their $5^{\textrm{th}}$ nearest neighbour will have low values of $\eta_{k}$.
 
\begin{figure*}
\includegraphics[width=0.9\textwidth]{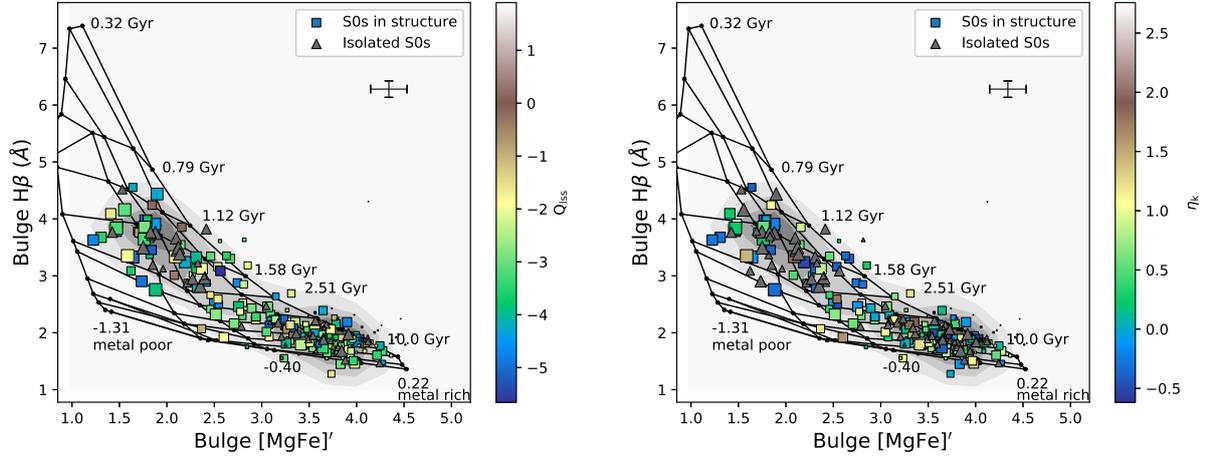}
\caption{The environments of MaNGA lenticular galaxies, as described by parameters introduced in \citet{Argudo-Fernandez15}. Bulge H$\beta$, [MgFe]$^{\prime}$, models, and errors are the same as for Figure~\ref{Hb_MgFe}a. S0s located in structure are denoted by squares, coloured by the environmental parameters Q$_{\textrm{lss}}$ (left) and $\eta_{k}$ (right). Squares are sized by their weighting within the MaNGA sample, and contours show this weighted distribution. Isolated S0s are denoted by grey triangles. There is no obvious environmental dependence on MaNGA lenticular stellar populations.}
\label{env}
\end{figure*}

 In Figure~\ref{env} we present bulge [MgFe]$^{\prime}$ and H$\beta$ measurements as a function of the environmental measures $Q_{\textrm{lss}}$ and $\eta_{k}$. Galaxies within structure are indicated with square coloured markers, and those deemed isolated by the method of \citet{Argudo-Fernandez15} are shown as grey triangles. We see a range of environments across the entire H$\beta$/[MgFe]$^{\prime}$ space, with both isolated S0s and those within denser regions or with tidal perturbations seen for all stellar masses, ages, and metallicities. From this we conclude that neither nearby neighbours perturbing S0s nor large-scale structure is singularly responsible for either population of S0s. The environmental parameters $Q_{\textrm{lss}}$ and $\eta_{k}$ are listed for each galaxy in Table~\ref{datatable}.
 
The caveat to this environmental investigation is that, while the MaNGA galaxy sample contains galaxies from a wide range of environments, there are no nearby cluster S0s currently included in the sample. While these environments are statistically rare, the lenticulars within them are well studied \citep[e.g.][]{Kuntschner00,Barway09, Bedregal11}. While somewhat concordant views on the star formation and stellar ages of bulges and disks of cluster S0s have been presented in the past \citep[e.g.][]{Silchenko06,Hudson10, Johnston14}, we cannot comment on whether this extreme environment will strongly affect stellar population trends seen throughout lower-density environments in the MaNGA sample.

\subsection{Formation/Quenching Pathways} %
\begin{figure*}
\includegraphics[width=0.8\textwidth]{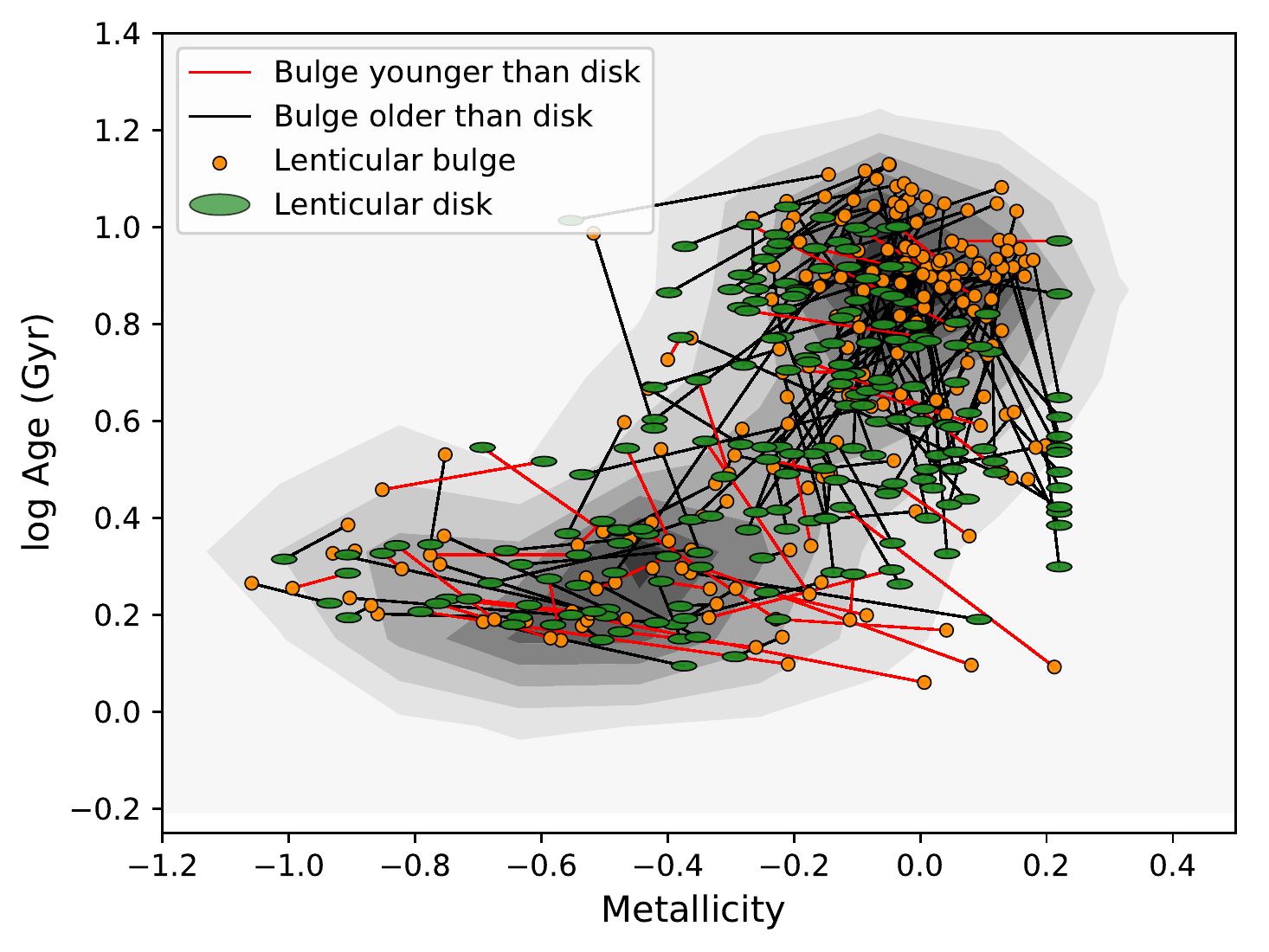} 
\caption{Bulge and disk stellar age and metallicity measurements for all MaNGA lenticulars. Here the bulge (orange) and disk (green) of a given galaxy are joined by a line. The line is black if the bulge of the galaxy is older than the disk, and red if the bulge is younger than the disk. Weighted contours are for the bulge regions. We see that the majority of galaxies with younger bulges lie in the low-mass, young, and metal-poor region of this diagram.}
\label{allthelines}
\end{figure*}

To investigate trends in bulge and disk age and metallicity further, we present the stellar population parameters for both galaxy components on the same plot in Figure~\ref{allthelines}. For each galaxy, the bulge age and metallicity measurement is joined to its corresponding disk value by a line. The line is black if the bulge is older than the disk and red if the bulge is younger than the disk. We immediately see the majority of high-mass, old, and metal-rich lenticulars possess bulges that are older than their disks, while the bulk of young, metal-poor and lower-mass lenticulars host bulges that are younger than their disks. The reversal of the older component between high-mass and some low-mass lenticulars gives further weight to a scenario in which two separate formation mechanisms are responsible for the overall lenticular population. Of the galaxies in the high-mass, metal-rich and older sample defined in Figure~\ref{Hb_MgFe}b, 84.7\% of galaxies possess bulges that are older than their disks. For the low-mass, metal-poor, and younger S0 sample, 50.8\% of galaxies possess bulges that are younger than their disks, though low-mass galaxies comprise 78.8\% of all galaxies with bulges younger than their disks in the entire S0 sample.

In Figure~\ref{allthelines} we see that the majority of old, metal-rich and massive lenticular galaxies possess bulges that are older than their disks. This, coupled with passive mid-IR colours, is consistent with the concept of morphological quenching. \citet{Martig09} explain that the growth of a galaxy's bulge (for example, through repeated mergers) will make the disk stable against fragmentation, quenching star formation. Indeed, the majority of high-mass lenticulars in this sample possess stellar populations $>5$ Gyr old. 
The older stellar population within the bulge implies the core quenched first, consistent with an inside-out quenching scenario \citep[e.g.][]{Mendel13, Tacchella15, Spindler17, Ellison18}. 

In contrast, many young, metal-poor, and low-mass lenticulars contain bulges that are younger than their disks. This is in line with simulations of spiral striping scenarios of \citet{Bekki11}. We postulate that while the higher-mass galaxies may have built up bulges large enough to undergo morphological quenching, this has not occurred in the low-mass galaxies. Instead, they may have undergone gas accretion processes, rejuvenating their bulges and causing a burst of recent (or current) star formation \citep[e.g.][in spiral galaxy bulges]{Thomas06}. This is in line with a `compaction' scenario, in which an episode of gas inflow is triggered by gas-rich (mostly minor) mergers \citet[e.g.][]{Johnston14}, \citet{Zolotov15} and \citet{Tacchella16}.
From Figure~\ref{allthelines}, we see the population of S0s with bulges younger than their disks has a much smaller variation in difference in bulge and disk stellar age when compared to the population of S0s with bulges older than their disks. This is consistent with a downsizing argument whereby more massive galaxies (those with bulges older than their disks) have formed their stars earlier and over a shorter timespan \citep[e.g.][]{Thomas05}. 
The majority of low-mass S0s also possess bulge S\'{e}rsic indices $<2$, indicating the population is dominated by pseudo-bulges \citep{Fisher08}. We expect these bulges will be preferentially formed via secular processes as opposed to violent, high-redshift mergers. This secular formation scenario for low-mass S0s supports findings such as those of \citet{Bellstedt17}, who find from simulations that low-mass S0s more closely resemble spiral progenitors than merger remnants.

We conclude that a scenario in which many low-mass S0s are formed by environmental stripping mechanisms, and high-mass systems are created by either morphological mechanisms or high-redshift dissipational processes, best fits the results of this study.

\section{Summary and Conclusions}
\label{conclusions}
To investigate the stellar populations of the bulge and disk regions of lenticular galaxies, we examine a carefully-selected sample of 279 S0 galaxies with a wide range of stellar masses from the MaNGA survey.
We measure H$\beta$, Mg$b$, Fe5270, and Fe5335 Lick indices from which we infer light-weighted stellar age, metallicity, and $\alpha$-enhancement parameters from both the bulge and disk regions of all galaxies in the lenticular sample. Despite some bulge light contamination within disk regions of high-mass S0s, we find:
\begin{itemize}
\item A bimodal age and metallicity distribution in both the bulge and disk regions of S0s. This bimodal distribution correlates best with stellar mass, such that low-mass lenticulars possess young and metal-poor stellar populations in both their bulges and disks, and high-mass lenticulars are old and metal-rich across the entire galaxy. From this we may infer that two separate formation sequences are responsible for the populations observed -- plausibly a stripped spiral scenario for the lower-mass galaxies, and mergers or other high-redshift dissipational processes for the higher-mass lenticulars. These two separate populations are not correlated with environment, however.

\item Despite differences in bulge and disk age between individual galaxies, the structural components of high-mass lenticulars are on average older and more metal-rich than their lower-mass counterparts. From this we conclude that galaxy bulges and disks evolve concurrently, with their formation histories and quenching pathways linked. 

\item When we examine the age differences between individual galaxies, we find that lower-mass lenticulars tend to have bulges that are marginally younger than their disks, whereas the opposite is true of high-mass galaxies. The high-mass galaxies can have quite an age difference between their old bulges and comparatively younger disks, but have mostly old stellar populations, from which we conclude high-mass lenticulars quenched inside out, perhaps through a morphological quenching process. 

\item Lower-mass galaxies are likely to have $n_{\textrm{b}}<$2, from which we imply either environmental mechanisms have stripped fuel for star formation, or the disk has simply run out of gas, fading the spiral arms. These low-mass galaxies that are more star forming may have undergone gas accretion processes that have rejuvenated their bulges into one last burst of more recent or even current star formation. These low-mass lenticulars may be undergoing or have undergone a compaction scenario.
\end{itemize}

Further work on spatially-resolved populations of lenticulars and separating components using bulge/disk decomposition on even larger samples (including denser cluster environments) will reveal whether the two-mechanism formation scenario correlates with stellar mass alone, and what we may infer about lenticular formation mechanisms.

\section{Acknowledgements}
The authors would like to gratefully acknowledge Eric Emsellem and Cheng Li for useful discussions on the subject, and the anonymous referee for useful comments that improved this paper.
AFM acknowledges support from STFC.
MAF is grateful for financial support from the CONICYT Astronomy Program CAS-CONICYT project No.\,CAS17002, sponsored by the Chinese Academy of Sciences (CAS), through a grant to the CAS South America Center for Astronomy (CASSACA) in Santiago, Chile.
Funding for the Sloan Digital Sky Survey IV has been provided by the Alfred P. Sloan Foundation, the U.S. Department of Energy Office of Science, and the Participating Institutions. SDSS-IV acknowledges
support and resources from the Center for High-Performance Computing at
the University of Utah. The SDSS web site is www.sdss.org.

SDSS-IV is managed by the Astrophysical Research Consortium for the 
Participating Institutions of the SDSS Collaboration including the 
Brazilian Participation Group, the Carnegie Institution for Science, 
Carnegie Mellon University, the Chilean Participation Group, the French Participation Group, Harvard-Smithsonian Center for Astrophysics, 
Instituto de Astrof\'isica de Canarias, The Johns Hopkins University, 
Kavli Institute for the Physics and Mathematics of the Universe (IPMU) / 
University of Tokyo, Lawrence Berkeley National Laboratory, 
Leibniz Institut f\"ur Astrophysik Potsdam (AIP),  
Max-Planck-Institut f\"ur Astronomie (MPIA Heidelberg), 
Max-Planck-Institut f\"ur Astrophysik (MPA Garching), 
Max-Planck-Institut f\"ur Extraterrestrische Physik (MPE), 
National Astronomical Observatories of China, New Mexico State University, 
New York University, University of Notre Dame, 
Observat\'ario Nacional / MCTI, The Ohio State University, 
Pennsylvania State University, Shanghai Astronomical Observatory, 
United Kingdom Participation Group,
Universidad Nacional Aut\'onoma de M\'exico, University of Arizona, 
University of Colorado Boulder, University of Oxford, University of Portsmouth, 
University of Utah, University of Virginia, University of Washington, University of Wisconsin, 
Vanderbilt University, and Yale University.
\appendix
\section{Data Tables}

\onecolumn
\begin{landscape}
\begin{footnotesize}

\begin{longtable}{l c c c c c c c c c c c c c c}
\caption{Physical properties and derived quantities of the lenticular galaxies used in this work. Galaxy positional coordinates, photometric-derived structural parameters, derived stellar population measurements, and environmental indicators are included in this table. We present the first ten lines as an illustrative example, and the rest will be available as an online table.}\\
  \label{datatable}\\
\hline
\hline
\textbf{MaNGA}   &   \textbf{Right}         &   \textbf{Declination}       & \textbf{Redshift} &   \textbf{Stellar Mass}    &  \textbf{b/a} & \textbf{B/T} & \textbf{R$_{b}$} &\textbf{n$_{b}$} &\textbf{Bulge}   &   \textbf{Disk}   &   \textbf{Bulge }   &   \textbf{Disk}   &   \textbf{Q$_\textbf{lss}$}  &      \textbf{$\eta_{\textbf{k}}$}     \\  
 \textbf{Plate-IFU}&\textbf{Ascension (deg)} &\textbf{(deg)} &  & \textbf{($\times 10^{10}~\textrm{M}_{\odot}$)}&  & & & &\textbf{$\log (\textrm{Age})$}  &\textbf{$\log (\textrm{Age})$}  &\textbf{[Z/H]} &\textbf{[Z/H]} &   &\\
 \hline
                 &   (1)    &   (1)    &   (2)   &   (2)      &   (3)   &   (3)      &   (3)    &   (3)   &             &           &             &            &   (4)      \\
\hline
  8078-3702    &     02:48:22   &  +00:59:11.3   &  0.0274   &  0.195  &  0.96   &  0.09  &  2.52    &  0.52   &  0.242  &  0.558  &   $-$0.175  &   $-$0.34  &   $-$2.325   &   $-$0.371     \\
  8077-9102    &     02:48:34   &   $-$00:32:59.5   &  0.0245   &  0.161  &  0.25   &  0.25  &  1.6     &  1.18   &  0.018  &   $-$0.016  &  \--  &  \--  &   $-$1.598   &  0.252  \\
  8077-3704    &     02:48:48   &   $-$00:06:33.1   &  0.0247   &  1.2  &  0.75   &  0.59  &  1.25    &  2.99   &  0.333  &  0.317  &   $-$0.206  &   $-$0.25  &   $-$4.698   &   $-$0.275     \\
  8081-1901    &     03:11:40   &  +00:46:54.2   &  0.027    &  0.334  &  0.26   &  0.37  &  2.03    &  1.63   &  0.234  &  0.184  &   $-$0.904  &   $-$0.418     &  Isolated   &  Isolated     \\
  8081-12701   &     03:12:47   &   $-$01:01:22.3   &  0.0815   &  18.1  &  0.89   &  0.47  &  2.87    &  2.39   &  0.482  &  0.422  &  0.144  &  0.22  &   $-$0.728   &  0.131  \\
  8081-3701    &     03:14:18   &   $-$00:36:34.9   &  0.1153   &  21.2  &  0.61   &  0.9  &  10.3    &  6.01   &  0.525  &   $-$1.1  &  0.22  &  0.22  &   $-$1.512   &  0.395  \\
  8082-3702    &     03:16:26   &  +00:19:17.9   &  0.0214   &  0.0888  &  0.25   &  0.04  &  0.7     &  1.76   &  0.53  &  0.345  &   $-$0.752  &   $-$0.776     &   $-$3.875   &   $-$0.308     \\
  8082-6101    &     03:20:35   &   $-$01:05:45.2   &  0.0214   &  0.0799  &  0.63   &  0.33  &  0.93    &  1.11   &  0.201  &  0.195  &   $-$0.859  &   $-$0.635     &   $-$0.475   &  2.758  \\
  8082-1902    &     03:20:39   &   $-$01:02:06.1   &  0.021    &  0.698  &  0.59   &  0.36  &  0.53    &  5.06   &  0.462  &  0.416  &   $-$0.178  &   $-$0.224     &   $-$1.307   &  2.451  \\
  8084-9101    &     03:22:51   &  +00:08:22.8   &  0.0228   &  0.173  &  0.38   &  0.62  &  2.29    &  1.92   &   $-$0.005  &  0.16  &  \--  &   $-$0.39  &  0.148    &  1.066  \\

  \hline
 \multicolumn{14}{l}{(1) J2000}\\
 \multicolumn{14}{l}{(2) Values obtained from the NASA Sloan Atlas.}\\
 \multicolumn{14}{l}{(3) Values obtained from the free $n_{\textrm{b}}$ model table of bulge and disk measurements of \citet{Simard11}.}\\
 \multicolumn{14}{l}{(4) Derived using the method of \citet{Argudo-Fernandez15}.}\\
  \hline
  \hline
  \end{longtable}
  \end{footnotesize}
 \end{landscape}
 
 \begin{footnotesize}
 \begin{longtable}{l c c c c c c c c}
\caption{Measured spectral indices for lenticular galaxies in this work. Here, we present the averaged bulge and disk index measurements in \AA ngstr\"{o}m, for each galaxy in the S0 sample. We present the first ten lines as an illustrative example, and the rest will be available as an online table.}\\
  \label{datatable2}\\
\hline
\hline
\textbf{MaNGA} & \textbf{Bulge}  & \textbf{Disk} &  \textbf{Bulge}  & \textbf{Disk}  & \textbf{Bulge}  & \textbf{Disk}  &  \textbf{Bulge}  & \textbf{Disk}  \\
\textbf{Plate-IFU} & \textbf{H$\beta$}  & \textbf{H$\beta$}  & \textbf{Mg$b$}  & \textbf{Mg$b$}  & \textbf{Fe5270}  & \textbf{Fe5270}  & \textbf{Fe5335}  & \textbf{Fe5335}  \\

 \hline
  8078-3702    &     3.161      &      2.315     &   2.288       &   2.53       &   3.065          &   3.08          &   2.163          &   2.231         \\
  8077-9102    &     4.092      &      4.315     &   1.43        &   1.616      &   1.319          &   1.925         &   1.563          &   2.004         \\
  8077-3704    &     2.877      &      2.929     &   2.724       &   2.745      &   2.806          &   2.509         &   2.649          &   2.709         \\
  8081-1901    &     3.765      &      3.793     &   1.696       &   1.992      &   1.147          &   1.854         &   1.326          &   1.836         \\
  8081-12701   &     2.42       &      2.565     &   4.953       &   8.053      &   2.73           &   3.081         &   2.728          &   2.656         \\
  8081-3701    &     2.314      &      \--     &   4.469       &   \--      &   3.839          &   \--         &   2.759          &   \--         \\
  8082-3702    &     2.757      &      3.27      &   1.684       &   1.741      &   2.066          &   2.025         &   2.197          &   1.581         \\
  8082-6101    &     3.914      &      3.734     &   1.027       &   1.516      &   2.582          &   2.15          &   1.0            &   1.588         \\
  8082-1902    &     2.527      &      2.632     &   2.849       &   2.844      &   2.828          &   2.688         &   2.562          &   2.289         \\
  8084-9101    &     4.243      &      3.513     &   1.592       &   1.643      &   2.373          &   2.353         &   1.534          &   2.174         \\

  \hline
  \hline
  \end{longtable}
  \end{footnotesize}
 
\twocolumn

    \bibliographystyle{mnras}
  \bibliography{MaNGAbib}
\end{document}